\documentclass[lettersize,journal]{IEEEtran}
\usepackage{amsmath,amsfonts}
\usepackage{algorithmic}
\usepackage{algorithm}
\usepackage{array}
\usepackage[caption=false,font=footnotesize,labelfont=sf,textfont=sf]{subfig}
\usepackage{textcomp}
\usepackage{stfloats}
\usepackage{url}
\usepackage{verbatim}
\usepackage{graphicx}
\usepackage{cite}
\usepackage{amssymb}
\usepackage{multirow}

\usepackage{algorithm}
\usepackage{algorithmic}

\renewcommand{\algorithmicrequire}{ \textbf{Input:}} 

\usepackage{cuted}

\usepackage{enumitem}

\newcommand*{\Scale} [2] [4]{\scalebox{#1}{$#2$}}%

\usepackage[backref]{hyperref} 
\hypersetup{ hidelinks}
\usepackage{xurl} 

\usepackage[justification=centering, font=small]{caption}

\DeclareMathSizes{10}{9}{7}{6}

\title{Personalized Playback Technology: How Short Video Services Create Excellent User Experience}

\author{
    Weihui Deng\thanks{All the authors are with Bytedance Inc.}
    , Zhiwei Fan\thanks{Sorted by alphabetical order of surnames. Weihui and He. L mainly contributed to user-item aware encoding (UIAE), Zhiwei and Chunyu focused on personalized streaming \& playback, Deliang led tech evolution direction from 2022, Yun, Zheng and Zhengyu mainly contributed to foresight based publishing, Shenglan contributed to all streaming, playback and encoding areas, Xiaocheng established the technical field from 2019,  Yiting and Bin contributed both on encoding and publishing. Additionally, Xiaocheng and Bin polished the whole text, Deliang and Zhen took a full internal review.}, Deliang Fu, Yun Gong, Shenglan Huang \\
     Xiaocheng Li, Zheng Li, Yiting Liao, He Liu \\
     Chunyu Qiao, Bin Wang,~\IEEEmembership{Member, IEEE}, Zhen Wang, Zhengyu Xiong \thanks{e-mails:
     \{dengweihui,fanzhiwei.rice, fudeliang, gongyun, huangshenglan, lixiaocheng.rd, lizheng.james, liaoyiting, he.liu.l, qiaochunyu, wangbin.hd, wangzhen3560, xiongzhengyu\}@bytedance.com}
}
\date{Oct 2024}

\begin{document}

\maketitle

\begin{abstract}
Short form video content has become increasingly popular recent years. Its concise yet engaging format aligns well with todays' fast-paced and on-the-go lifestyles, making it a dominating trend. As one of the front runners, ByteDance has been highly successful in delivering a one-of-a-kind short video experience and attracting billions of users worldwide. One key contributing factor is its advanced E2E personalized short video playback technology, where we pioneered and developed the new technical field over the past five years to optimize user experience. This paper introduces the major concepts and methodologies of this technology that distinguish it from traditional multimedia works. More details, including goal setting, iterative process, modeling, are also provided to encourage deeper research. 
\end{abstract}

\section{Introduction}
\IEEEPARstart{V}{ideo} has taken up 65\%~\cite{Sandvine2023} of global traffic, and short videos~\cite{VideoClip} have rapidly gained worldwide popularity due to high information density and fragmented consumption patterns. Traditionally, the attentions in multimedia industry has primarily concentrated on below directions: 
\begin{itemize}[leftmargin=*]
    \item \textbf{Codec, pre- and post-processing algorithms.} E.g, video codecs like MPEG-2\cite{MPEG-2} , RV\cite{RV}, AVC\cite{AVC}, HEVC\cite{HEVC}, VVC\cite{VVC}, VP9\cite{VP9}, AV1\cite{AV1}, audio codecs like MP3\cite{MP3}, AAC\cite{AAC}, HEAACv1/v2\cite{HEAAC}, OGG\cite{Ogg}, image codecs like Jpeg\cite{JPEG}, WebP\cite{WebP}, Heif\cite{HEIF}, pre- and post-processing such as super-resolution\cite{SRCNN, yang2010image, ledig2017photo}, de-noising\cite{tassano2020fastdvdnet, xu2020learning}, color enhancement\cite{kim2021representative, xue2019video} and ROI region analysis\cite{cai2019end, perugachi2022region}. 
    \item \textbf{Streaming protocols.} Some widely applied protocols include RTP/RTSP\cite{RTSP}, MPEG2-TS\cite{MPEG2-TS}, RTMP\cite{RTMP}, HLS\cite{HLS}, DASH\cite{DASH}, SRT\cite{SRT}, RTC\cite{RTC}. 
    
    \item \textbf{Multimedia frameworks.} Examples include DirectShow\cite{DirectShow}, Helix\cite{Helix}, MediaFramework\cite{MediaFramework}, AVFoundation\cite{AVFoundation}, FFMpeg\cite{ffmpeg}, VideoLAN\cite{VideoLAN}, Gstreamer\cite{GStreamer}. 

    \item \textbf{Video cloud.} E2E function set based on elastic computing platforms\cite{VideoEncodingNetflix, Borg2015, Hulu}, including metadata, upload, transcoding, delivery, CDN, playback SDK and analytics dashboard, which can be quickly integrated as on public cloud (like Brightcove, AWS and Google Cloud).    
\end{itemize}

In addition, other important works, such as signal modulation, DRM, custormized chipset, etc, which are interconnected with the above directions, influencing each other's trajectories, will not be elaborated. 

In direction of the control and decision optimization, industry and academia have made numerous attempts (e.g ABR~\cite{FESTIVE, bola, pensieve, oboe, Fugu, rl-abr}, CAE~\cite{CAE-Netflix, mico2023per, CAE-brightcove, per-sceneEncoding}). Meta shared its own encoding pipeline \cite{Chess, metaEncoding}. YouTube also conducted extensive research \cite{mondal2017candid, anorga2018analysis, wamser2016modeling}. However, they are localized optimization and unsystematic, and cannot adequately address the issues encountered by short videos. 

During our iteration, we face unprecedented challenges:
\begin{itemize}[leftmargin=*]
    \item \textbf{Quantity of videos and their diversity.} Short video app has 100 mil level UGC publishes everyday, meanwhile video contents and qualities vary greatly. Many known solutions and their principles, are carried out on a much smaller scale. 
    \item \textbf{Number of users and their diversity.} Billions of users watch short videos every day, and their habbits and preferences vary greatly. Although signal number is huge, valued information inside is sparse. Impact of multimedia treatment such as image quality and fluency is relatively indirect, effect is easily concealed. 
    \item \textbf{Revenue and cost constraints with complex business forms.} Short video applications currently host extremely complex businesses, including advertising, e-commerce, live broadcast rewards, takeaway services, search, etc. At same time, cost of video processing, storage and distribution are extremely high. Technology should balance user experience, business and costs. 
\end{itemize}

By introducing personalized technologies that are developing rapidly in fields of recommendation, advertising, user growth, etc., and combining them with traditional multimedia technologies, 1st time in industry, we had built a complete system since 2019, bringing a fundamental paradigm shift. Our system has achieved excellent results both in internal evaluations and competition. 

\section{Overview} \label{Overview}

\subsection{Methodology}
Warren E. Buffett~\cite{BuffettLetter} has argued that value of enterprise, and even intrinsic value of any asset $T$, is essentially present value of its future cash flows $D_t$ discounted at an appropriate interest rate of $r$, which can be written as  \eqref{equ:Buffett}: 
\begin{equation}
\begin{footnotesize}
\begin{aligned}
    T =  \frac{D_1}{1+r}+\frac{D_2}{\left  ( 1+r \right ) ^2} +\cdots +\frac{D_n}{\left  ( 1+r \right ) ^n}  =\sum\limits_{t=1}^{n}  \frac{D_t}{\left  ( 1+r \right ) ^t} \label{equ:Buffett}
\end{aligned}
\end{footnotesize}
\end{equation}

According to this assertion, assuming revenue from a user $u$ at a time $t$ is originally $D^u_t$, and impact of user $u$'s change through a certain technical modification is $\Delta d^u_t$, that is, business increase from $u$ has been changed from $D_t^u$ to $D_t^u+\Delta d^u_t$ after the technology launched, then gain of this technology for this certain $u$ can be expressed as  \eqref{equ:Tu}:
\begin{equation}
\begin{footnotesize}
\begin{aligned}
    \Delta T^u  = \frac{\Delta d^u_1}{1+r} + \frac{\Delta d^u_2}{ (1+r)^2} + \cdots + \frac{\Delta d^u_n}{ (1+r)^n}
    = \sum\limits_{t=1}^{n} \frac{\Delta d^u_t}{ (1+r)^t} \label{equ:Tu}
\end{aligned}
\end{footnotesize}
\end{equation}

Then, a to-C company like short video platform, its value of entire company can be expressed as the sum of discounted revenue of all potential users, work in this area is to maximize the positive sum of impact of them. 

Note, it is just a conceptual description, we should evaluate impact on users for every changes. Besides, there are some obvious constraints such as restriction of technology itself and investable quota, which means to consider marginal effect, fundraising cycle and market ceiling. 

We can see that user is the core, the concept of \textbf{personalized playback technology} is to re-evaluate and selectively use various technologies in area, to provide overall optimal effect for user preferences, and comprehensively considering all dimensions such as content, context, device and network. Correspondingly, traditional practice does not pay enough attention to differentiated demands, and often focuses on solidified framework and capabilities, or trap into local better solutions, which lose huge improvement room with big scale. 

\begin{figure}
    \centering
    \includegraphics [width=0.75\linewidth]{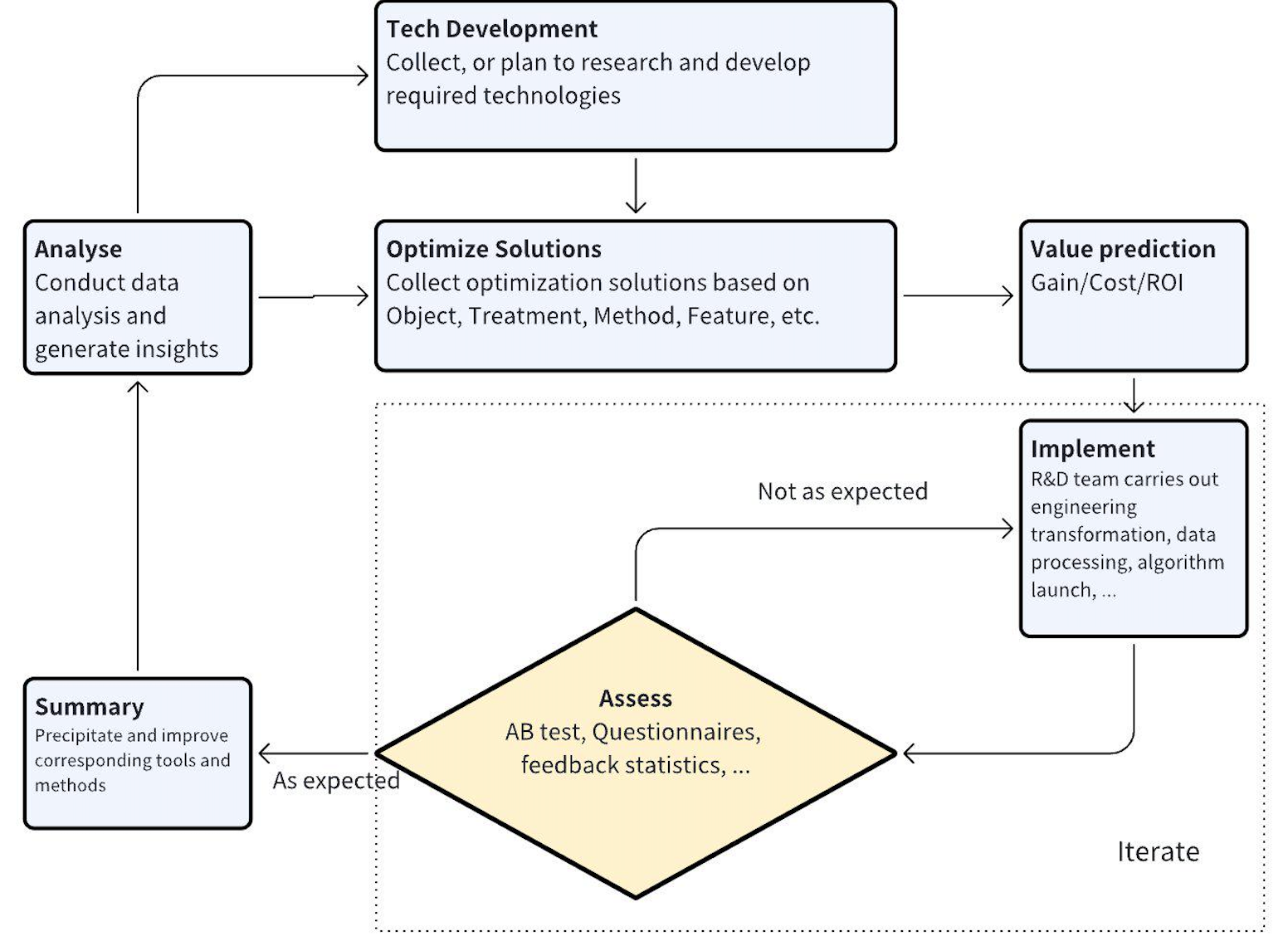}
    \caption{Common Workflow of Personalized Playback Technology} 
    \label{fig:workflow_psp}
\end{figure}

Thus, we have conducted the workflow, refer to Fig.~\ref{fig:workflow_psp}, which is similar to other personalized tech fields such as recommendation, will evaluate its user impact for each iteration. This is not a very new idea, but we place this approach at the core to drive traditional work such as codec, protocols, frameworks, and cloud, etc. 

\subsection{Goal Definition} \label{GoalDefinition}

From a general point of view, to-C social media enterprises widely use $LT$ (Life Time) to represent the activity of users over time, and $ARPU$  (Average Revenue Per User) to represent average revenue per user. For convenience of discussion below, let's simplify it, set overall goal of short video platform as $LTV$ (Life Time Value), see \eqref{equ:LTV}. 
\begin{equation}
\begin{footnotesize}
    LTV=LT \times ARPU \label{equ:LTV}
    \end{footnotesize}
\end{equation}

Changes in behavior metrics, such as play time and publishing quantities, actually reflect whether users are satisfied, can be converted into $LT$, and all revenue metrics can be converted into $ARPU$. Therefore, we can set an indicator $Profit$ to represent what technology brings, shown in \eqref{equ:ProfitDefinition} : 
\begin{equation} \label{equ:ProfitDefinition}
\begin{footnotesize}
        \begin{aligned}
      Pro&fit = \Delta LTV - \Delta COST   \\ 
     &=  (LT + \Delta LT) \times  (ARPU + \Delta ARPU)- LTV - \Delta COST  \\  
    & \approx  LT \times \Delta ARPU + \Delta LT \times ARPU  - \Delta COST  \\
   &  s.t. \ ROI = Profit  / \Delta COST  > \gamma
\end{aligned}
 \end{footnotesize}
\end{equation}

Here $\gamma$ denotes ROI coefficient. 

In multimedia area~\cite{streamingMetrics}, there are multiple $QoS$ indicators such as EBVS (Exist Before Video Start), VST (Video Startup Time), and Rebuffering Ratio. However, it is not enough to describe users, nor can fully reflect user state changes during playback, so we set up comprehensive indicators called $QoP$ (Quality of Performance) to represent multi-dimensional and complete performance experience. See Table~\ref{table:QoP} for partial of detailed definition. $QoP$ is an ideal all-encompassing description. We also conduct a detailed analysis to estimate the impact of each metric on business. 

\begin{table*} 
    \caption{
    PARTIAL OF $QOP$ METRIC DEFINITION
    }
    \label{table:QoP}
    \centering
    \begin{tabular}{p{4cm}p{7cm}p{3cm}}
    \hline
        \textbf{Metrics (Per 1\% Change)} & \textbf{Definition} & \textbf{LT Impact Magnitude}  \\
    \hline
    Power Consumption Average & Speed at which device consumes & 0.027\%+ \\
    Storage Average & Storaged used percentage on device  & 0.013\%+ \\
    First Feed Duration PCT50 & 50th percentile of loading time required to open 1st video  & 0.005-0.007\% \\
    First Frame Duration PCT50 / Average & Time to load 1st frame of video  PCT50/Avg \newline   (in feed, will degenerate to VST in \newline on-demand    playback scenario)  & 0.023\%+ \\
    Frame Drop Rate & 1-Playback frame rate/Video frame rate  & Not Available \\
    ANR/Crash User Rate & Failure to respond to user input or application crash & 0.0053-0.0062\%\\
    Temperature & Device heating temperature & 0.183\%+\\
    CPU Usage & CPU usage percentage average & 0.005-0.023\%\\
    OOM Rate & Out of memory times on device & 0.0009\%+\\
    Publish Success Ratio & Number of publish success on server / total submissions & Not Available\\
    FPS & Device refresh rate & 0.021\%+\\
    Rebuffering Ratio, Rebuffering Duration per vv & Total rebuffer times / video view, total rebuffer duration / total playback duration & 0.015\%+\\
    Memory Usage Rate & Device memory usage percentage average & 0.004\%+\\
    Traffic & Amount of traffic consumed by device & Not Available\\
    \hline
    \end{tabular}

\end{table*}

Note not all $QoP$ indicators are in table. Due to business stage, the indicators covered will be modified, and for each absolute value, impact may also change significantly. 

In evaluation process, we will calculate  \eqref{equ:ProfitDefinition}  for each project, and set different $\gamma$ to adjust cost values according to scenario and goals, and make it meet the launch guidelines: (1) $Profit > 0$, (2) attributable, observations of $QoP$ can give an explanation of impact that meets expectations. 

Evaluation methods are mainly based on online AB testing, questionnaires, feedback statistics, AA comparison, monitoring, attribution analysis, etc.

Based on above principles, we can avoid some newer but impractical technologies and build a solution far more complex than normal. 

\subsection{E2E Framework}
Traditionally, E2E multimedia technology divides pipeline into multiple components, including publishing, transcoding, delivery, scheduling, and playback\cite{AWSMedia}. 

\begin{figure}
    \centering
    \includegraphics [width=0.95\linewidth]{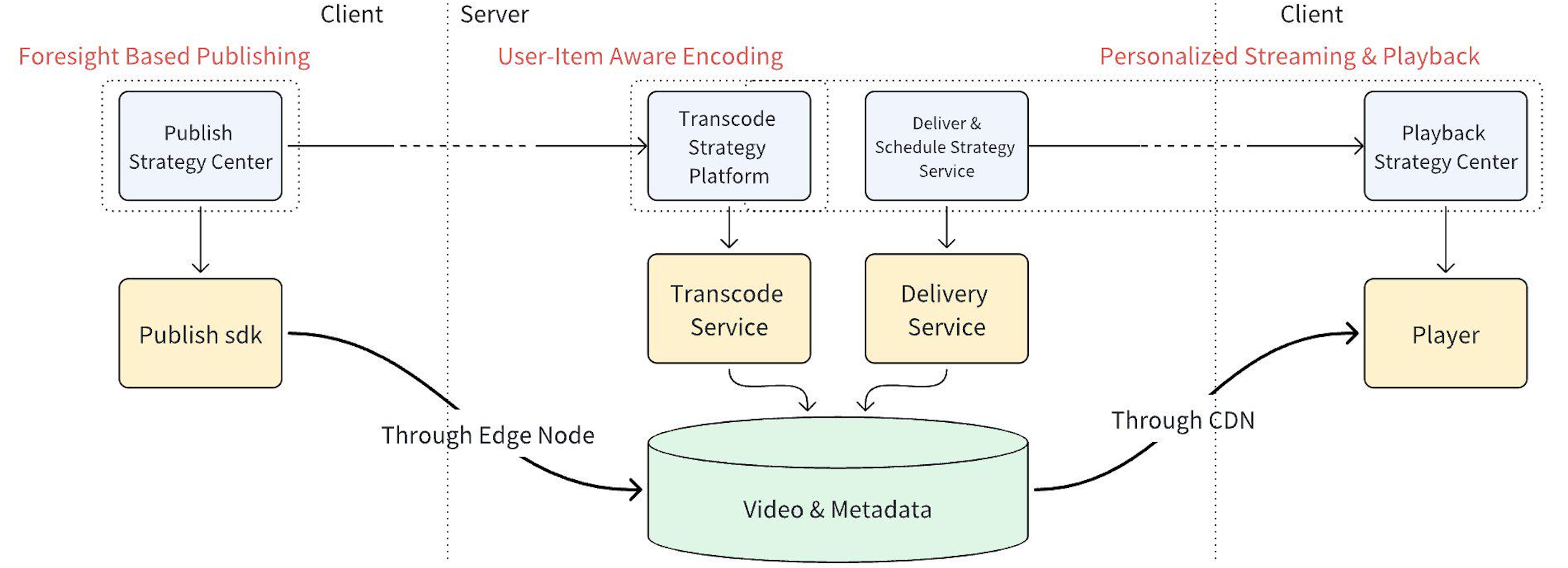}
    \caption{Conceptual architecture of personalized playback technology}
    \label{fig:framework_psp}
\end{figure}

In our system, special emphasis will be placed on cybernetics, as well as actual user impact, we believe that each sub-domain has co-ordinate space with others. Although the complex pipeline prevents direct overall modeling, each sub-domain still needs to deeply understand other segments to approximate best effect, so a conceptual division is as Fig.~\ref{fig:framework_psp}. 

As the architecture shown in diagram, strategies of delivery, scheduling, transmission, download, and playback should be considered as a whole, called Personalized Streaming \& Playback, which will be discussed in chapter~\ref{sec:PersonalizedStreamingPlayback}. Transcoding  strategy is named User-Item Aware Encoding, in Chapter~\ref{sec:UIAE}. Encoding and upload strategy associated with publishing is named Foresight Based Publishing, in chapter~\ref{sec:FP}. 

Personalization hints strategic control, which usually includes following perspectives: 
\begin{itemize} [leftmargin=*]
    \item \textbf{User:} Strategy for specific user should be individual, it can be subdivided into publishers, consumers, and even publisher $\times$ consumer combination. 
    \item \textbf{Item:} Differentiates each item (video, or other genres like image) to provide effects in user $\times$ item dimension. 
    \item \textbf{Context:} It can be user scenarios, personal state, environment, etc, which has differentiated in dimensions of user $\times$ context, item $\times$ context, or user $\times$ item $\times$ context, etc. 
\end{itemize}

For various treatment points for strategy (there are dozens or even hundreds of treatments in each sub-domain in Fig.~\ref{fig:framework_psp}, some of which will be introduced in following sections), level of optimization can be divided as follows: 
\begin{itemize}  [leftmargin=*]
    \item \textbf{Rule-based level:} Uses specific strategy for a group of users, items, or under special context, based on simple, explicit and limited rules. 
    \item \textbf{Portrait based level:} Portraits can be generated through data analysis, data mining, or model filtering for users and items, strategies designed for various groups. 
    \item \textbf{Individual level:} Strategy are used for a single user, item, considering its context, usually machine learning and deep models are required. 
\end{itemize}

There may be interlacing between multiple levels, for example, an portrait can not only be used separately, but can still be as input features processed by individual level strategy. 

\begin{figure}
    \centering
    \includegraphics [width=0.7\linewidth]{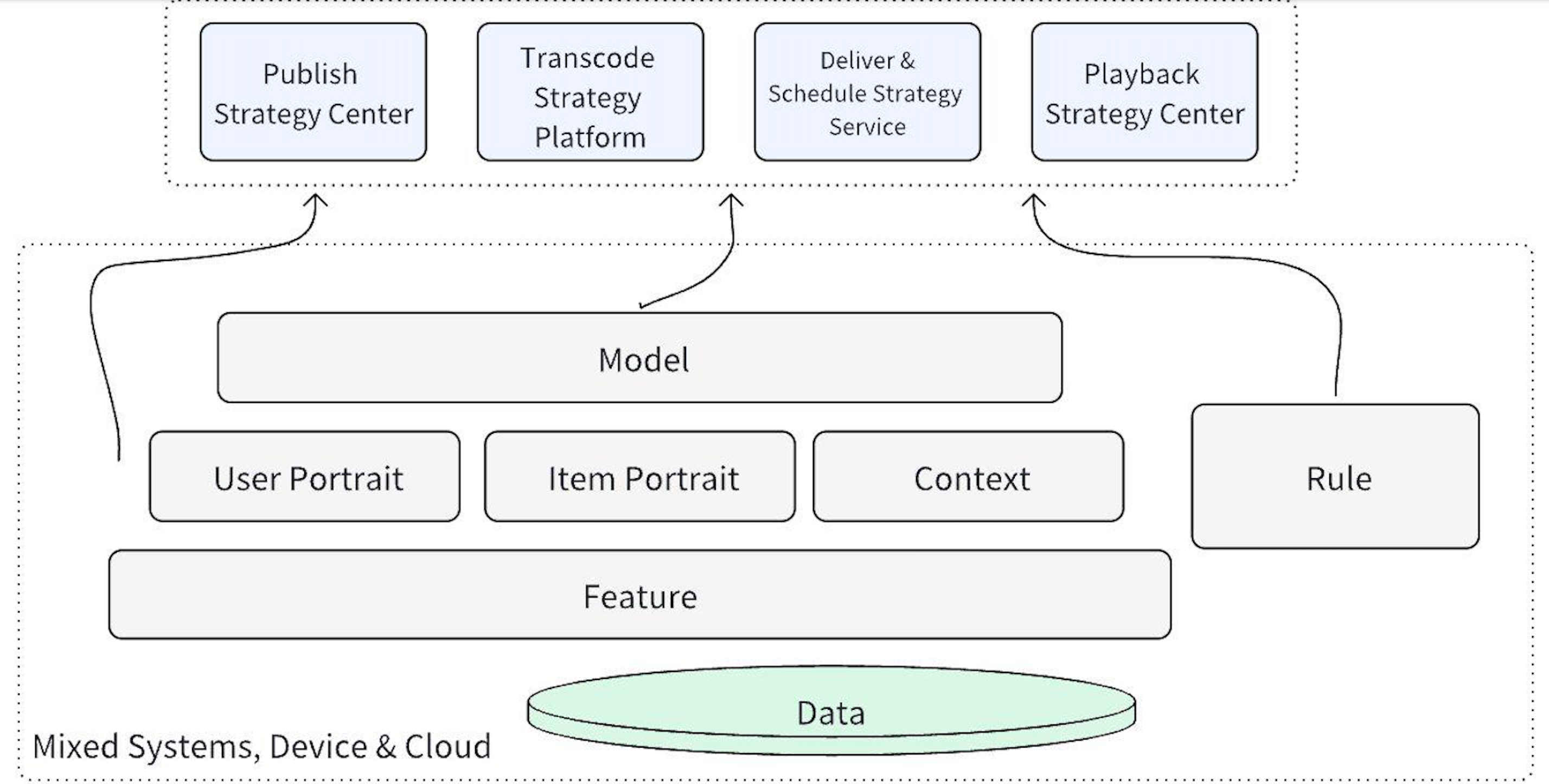}
    \caption{Conceptual Systems for Personalized Playback Technology}
    \label{fig:psp_high_level}
\end{figure}

The system (illustrated in Fig.~\ref{fig:psp_high_level}) can be implemented by raw data, rules, features, portraits, models (or algorithm modules), etc., and has corresponding capabilities on both server and client side  to support all strategies on decision points. 

\section{Personalized Streaming \& Playback} \label{sec:PersonalizedStreamingPlayback}
\begin{figure}[!h]
    \centering
    \includegraphics [width=0.95\linewidth]{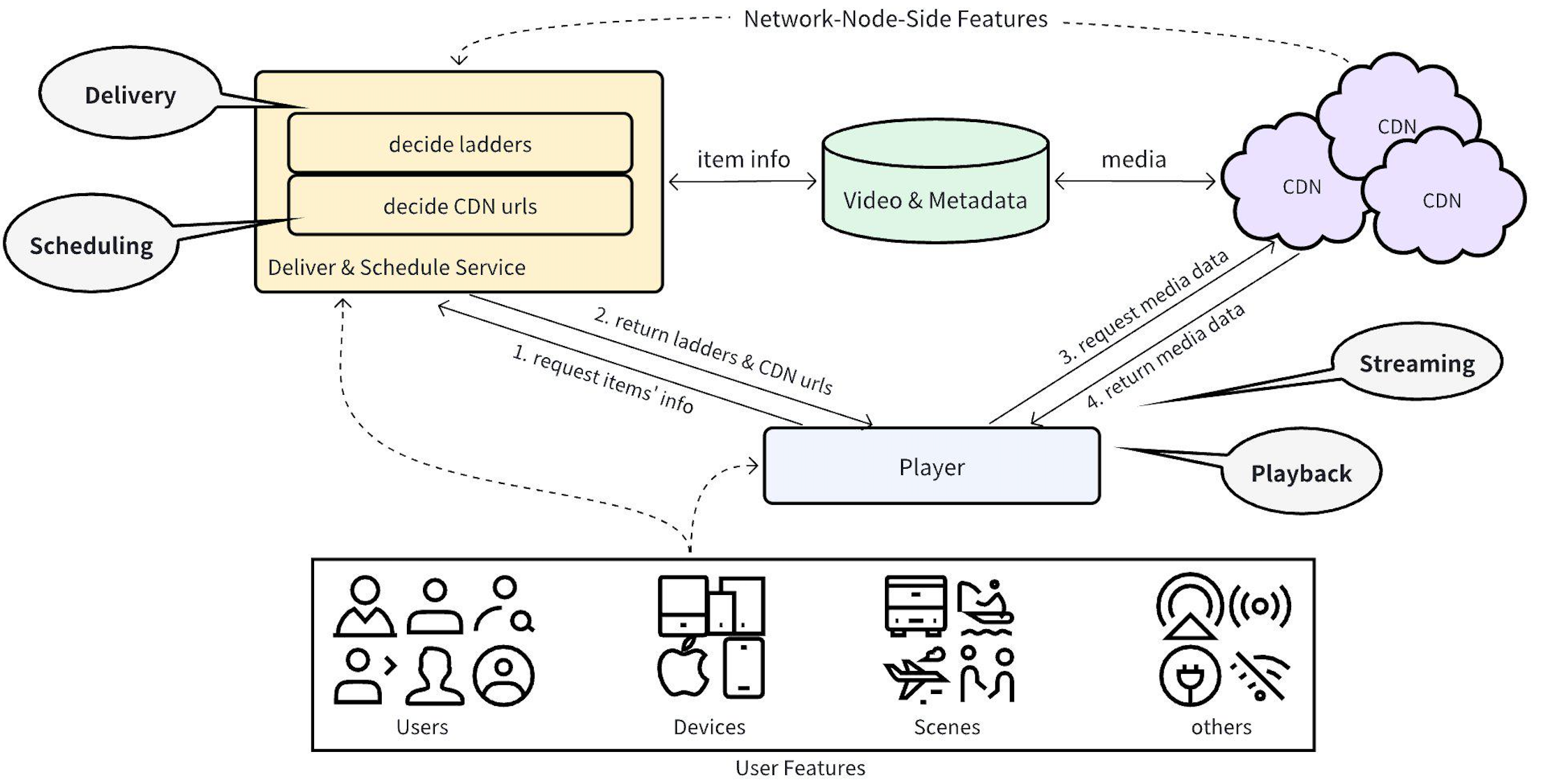}
    \caption{Diagram of Personalized Streaming \& Playback Module}
    \label{fig:diagram_psp}
\end{figure}

The entire playback service can be regarded as comprising three stages: Delivery, Scheduling, Streaming \& Playback. By performing algorithmic optimization of those stages based on available ladders of items, it effectively enhances playback experience, reduces cost, and ultimately improves overall business profitability. Diagram shown in Fig.~\ref{fig:diagram_psp}.

Optimization objective of entire streaming \& playback process described above can be expressed as  \eqref{equ:PsPObjective}: 
\begin{equation}
    \Scale [0.9]{
    \arg \max_{\pi (A|S)}\sum_{u\in U}LTV^{itl}_{u}\left  ( QoP^{itl}_{u}\left  ( \pi,s \right )| s \right ) - COST\left (\pi,s \right) \label{equ:PsPObjective}
    }
\end{equation} 
In above formula, 
\begin{itemize} [leftmargin=*]
    \item $S$ represents all relevant states, User, Item, and Context. 
    \item $A=<A_{deliver},A_{scheduler},A_{client}>$ represents all modules actions set, delivery, scheduling, streaming and playback. 
\item $\pi (A|S)$ is strategy for choosing action $a$ given state $s$.
\item  $QoP^{itl}_{u}\left (\pi,s \right)$ represents expectation of performance experience $QoP$ while $user$ play item list $itl$ , while execute actions according to strategy $\pi$, under overall state $s$. 
\item $itl$ includes specific video content and ladders. Its content is determined by recommendation algorithm and user selection,  ladders are determined by transcoding service. 
\item $LTV^{itl}_{u}\left (QoP|s \right)$ indicates expectation of business value from $u$ viewing $itl$ with performance experience of $QoP$ under overall state $s$. Therefore, sum of business value to all users, $\sum\limits_{u\in U}LTV_{u}^{itl}$ is total value of the product. 
\item $COST\left (\pi,s\right)$ is all costs while executing actions according to strategy $\pi$ under $s$, and in some problems of this chapter, total cost $COST$ is equal to all costs per $u$, we only need to focus on impact of single user or each action played, however, in other cases there is clear constraint that require concern about the sum. 
\end{itemize}

Optimization for streaming \& playback is to find optimal strategy $\pi$, and thinking of executability, it is necessary to ensure that iteration of $\pi$ can increase $Profit$ in a continuous and greedy way in each cycle. 
\subsection{Personalized Streaming \& Playback} \label{subsec:PsP}
According to characteristics of short video applications, feed playback occupies major use time. A single feed request retrieves a sorted item list for sequential watch. Player should anticipate viewing behavior by pre-downloading segments of each video that are most likely to be watched and preparing necessary components in advance. It guarantees startup speed, smoothness, and video quality in whole list. Other words, the optimization objective is no longer effect of a single playback action but overall experience throughout entire session. 

\noindent \textbf{Method}

For optimization of download and playback strategy, industry generally takes $QoS$ or $QOE$ which is indicated by weighted sum of some $QoS$ metrics \cite{pensieve, oboe, Fugu, beyondqoe, engagementinstreaming}, historical work is usually focused on download-related actions and cache control to achieve comprehensive optimization of fluency, quality and cost. However, it deviates greatly from user experience and business goal, video and contextual information are underutilized, at same time, process such as decoding and rendering is not controlled coordinately, further, joint optimization of playback and download is also lacking. 

Our approach expects to model directly with business goals, and since global constraints can be ignored approximately on client side, we can aim to maximize $Profit$ per user:
\begin{equation} 
    \Scale [0.9]{
    \arg \max_{\pi_{u} ( A_{client}|S_{client})}Profit^{itl}_{u}\left  (QoP^{itl}_{u}\left  ( \pi_{u},s_{u} \right )| s_{u} \right ) \label{equ:profit}
    }
\end{equation} 
\begin{itemize} [leftmargin=*]
    \item $A_{client}$ represents all actions executed on client side, including super-resolution, pre-render, pre-download, download size, etc. Since the choice can affect multiple dimensions, the corresponding action space dimension will be extremely high. It is necessary to divide and conquer, but is also important to coordinate in key dimensions.

    \item $S_{client}$ represents collections of user states such as device score, cache, profile, video states such as bitrate, video quality, predicted play time, etc., and context states such as app page, network environment, etc. 
    \item $s_{u}\in S_{client}$ indicates states of $u$ states. 
    \item $Profit_{u}^{itl}$ represents gain from $u$ while consuming $itl$. 
    \item $\pi_{u} (A_{client}|S_{client})$ is strategy to choose actions under $u$ states, for improving $Profit_{u}^{itl}$ . There are two paths:
    \begin{itemize} [leftmargin=*]
        \item Analogous to traditional idea of keeping most aspects of $QoS$ less degraded and significantly optimizing one or more aspects to leverage overall improvement, we upgrade traditional $QoS$ to $QoP$, that is, construct mapping of $QoS\to Profit$ as $QoP\to Profit$, introduce features to optimize $QoP$, so as to improve $Profit_{u}^{itl}$. 
        \item Personalized optimization method considering user preferences can be summarized as, discover performance preferences of users and scenarios in various dimensions$\longrightarrow$optimizing modeling of $QoP_{u}^{itl}\to Profit_{u}^{itl}$ $\longrightarrow$ iterate strategy $\longrightarrow$ obtain benefits. 
    \end{itemize}
\end{itemize}
\noindent \textbf{Action Space}

Streaming \& playback action space on device $A_{client}$ can be roughly divided into download and playback control actions. 

In our applications, it is necessary to ensure complete collection of actions, that is, to consider all factors (actions) that have certain impact on our objective. A feasible way to sort out is to build a matrix for all modules (components or code blocks within components) and dimensions like CPU, GPU, IO, Memory, Disk, Socket and other resources,  traverse  matrix elements, evaluate whether there are multiple implementation paths or wider parameter ranges, and estimate their impact on $QoP$. Those with highest impact will be put into decision-making pool. Look at formula ~\eqref{equ:action_matrix}, 
\begin{itemize} [leftmargin=*]
    \item $a_{ (i,k),j}$ represents usage of the $j$ resource by the $k$ implementation method of a module $i$.
    \item $QoP_{ (i,k),j}$ represents $QoP$ impact brought by $a_{ (i,k),j}$. 
    \item $\top_k$ represents the top K $QoP_{ (i,k),j}$ with greatest impact. 
    \item $A_{top}$ is the list of  ( (module, implementation), resource) pairs corresponding to $\top_k$, that is, action we expect to make strategic decisions. 
\end{itemize}

\begin{strip}
    \begin{footnotesize}
\begin{equation} \label{equ:action_matrix}
    \begin{aligned}
         A_{top}\! =\!\mathop{\arg}\limits_{QoP}\!\top_k\!(\!
        \left [
    \begin{matrix}
    a_{ (0,0),0} \! & a_{ (0,0),1} \!& \cdots \!& a_{ (0,0),N-1} \\
    a_{ (0,1),0} \!& a_{ (0,1),1} \!& \cdots \!& a_{ (0,1),N-1} \\
    \vdots & \vdots & \vdots & \vdots \\
    a_{ (i,k),j}\!&\!a_{ (i,k),j} \!&\!\cdots\!& a_{ (i,k),N-1} \\
    \!\vdots & \!\vdots & \!\vdots & \!\vdots \\
    a_{ (M-1,k),0}\!&\!a_{ (M-1,k),1}\!&\!\cdots\!&\!a_{ (M-1,k),N-1} \\
    \end{matrix}\!
    \!\right]\!\stackrel{eval}\Longrightarrow \!\left[\!
    \begin{matrix}\!
    QoP_{ (0,0),0} & QoP_{ (0,0),1} & \cdots & QoP_{ (0,0),N-1} \\
    QoP_{ (0,1),0} & QoP_{ (0,1),1} & \cdots & QoP_{ (0,2),N-1} \\
    \vdots & \vdots & \vdots & \vdots \\
    QoP_{ (i,k),j}\!&\!QoP_{ (i,k),j} & \!\cdots & QoP_{ (i,k),N-1} \\
    \vdots & \vdots & \vdots & \vdots \\
    QoP_{(M-1,k),0}\! &\! QoP_{ (M-1,k),1}\! &\!\cdots\!&\!QoP_{ (M-1,k),N-1} \\
    \end{matrix}\!
    \right]\!
    |k)
\end{aligned}
\end{equation}    
\end{footnotesize}
\end{strip}

\noindent \textbf{State Space} 

States can be considered as features required by strategy, larger state space means higher ceiling. In order to achieve personalized modeling, states of all levels such as rule-based, portrait-based, and individual should be systematically considered when constructing state space, and constructed from perspectives of user, item, and context. 

Basic feature construction needs to meet correlation, further, real-time features that can fully identify performance status should be constructed for aspects of each $QoP$ indicator, as well as predictive features for future performance~\cite{wang2024application}. 

There are multiple ways to build portraits. Firstly, it can be objective-driven, build corresponding portraits based on $QoP$ indicators to optimize, such as portraits of fluency-sensitive, video quality-sensitive, traffic-sensitive, storage-sensitive, etc. Secondly, prior information can be used for categories and levels of users. Thirdly, any attributes that can be divided from the perspectives of user, item, context and have a correlation with $QoP$ changes are available. 

Taking video quality sensitive portrait as an example, we use uplift model to predict quantile regression of treatment or control playback time, and divides users into different groups $U_g$ based on uplift-value. Obviously, this portrait will provide key information related to the balance between fluency and video quality, see (\ref{equ:uplift}), in which, $Y_t$ is value of treatment strategy, $Y_0$ is value of control strategy, $w$ is actual uplift model parameter, $U_g$ is user group aggregated based on uplift-value. 
\begin{equation}
\begin{footnotesize}
    \begin{aligned}  \label{equ:uplift}
        & \mathrm{uplift} =  (Y_t (S,w)-Y_0 (S,w))  \\
        & U_g = \begin{cases}
        U_1 & \text{uplift  } \leq thr_1,\\
        U_2  & \text{uplift  } > thr_1 \& \ \text{uplift  } \le thr_2,\\
        \vdots\\
        U_n  & \text{uplift  } > thr_{n-1} \& \ \text{uplift  } \le thr_n 
        \end{cases}
         \\
    \end{aligned}
    \end{footnotesize}
\end{equation} 

Construction of individual-level features usually requires to be classification, regression value or probability distribution through calculation or prediction. For features like a certain user-item playback duration, historically, some work \cite{pdas, Duasvs, LiveClip} has predicted it, either based on results of user's past few plays, or using distribution of video watched by all users instead of specific user's viewing time distribution. However, in our applications, video consumption presents a geometric distribution and life cycle is very short. Such practices are too rough and do not fully consider personalization. 

A simplified and feasible approach is to establish distribution of playback duration from the source duration, user, and video dimensions, and use weighted fitting of parameters, see following formula \eqref{equ:duration}, to merge them into a comprehensive duration distribution. 
\begin{equation}
    \begin{aligned} \label{equ:duration}
    playtime& (u,i) =  \alpha_1 \times playtime_{bucket} (i\to bucket) \\
    & +\alpha_2 \times playtime_{item} (i)   + \alpha_3 \times playtime_{user} (u) 
    \end{aligned}
\end{equation}
\begin{itemize} [leftmargin=*]
    \item $playtime (u,i)$ indicates estimated playtime of $u$ on $i$. 
    \item  $playtime_{bucket}$ is playtime distribution modeled based on duration of source video, which is divided into buckets, and calculated for each bucket. $i\to bucket$ indicates mapping playtime of video to corresponding duration bucket. 
    \item $playtime_{item}$ is estimated playtime distribution based on historical playtime data of video $i$. 
    \item  $playtime_{user} (u)$ is playtime distribution of $u$. 
\end{itemize}

Similarly, we can also use complex deep models to make more accurate predictions, see \eqref{equ:duration_nn}: 
\begin{equation}
playtime (u,i) = NN (user,item, playback\_seqs) \label{equ:duration_nn}
\end{equation}

Here, $playback\_seqs$ is user's viewing sequence and corresponding item information in a session. 

\noindent \textbf{Optimization Example}

Deciding which ladder to download at a specific time, is a common sub-problem of streaming area. The related problem in short video feed playback scenario is download control strategy \cite{pdas, Duasvs, LiveClip}. Optimizing these two control algorithms can be regarded as a decision for two (groups of) actions. Since scope of action space and state space to be optimized in our work is much larger than before, and tools used are also highly extensive, here we only use ladder selection and download control as examples for introduction. 

Traditional ladder selection and download control algorithms generally set their goal to maximize $QoE$ represented by $QoS$ \cite{mpc, pensieve, Fugu}, see  \eqref{equ:QoE}: 
\begin{equation} 
\begin{footnotesize}
    \begin{aligned}
    \max QoE-\gamma Cost & = Quality-\alpha BlockDur \\
    & -\beta QualitySwitch -\gamma Cost \label{equ:QoE}
\end{aligned}
\end{footnotesize}
\end{equation}

Here $Quality$ indicates quality of video, which is generally expressed by bitrate or other related attributes, $BlockDur$ indicates duration of rebuffering, $QualitySwitch$ indicates degree of ladder switching, $Cost$ indicates bandwidth cost, and $\alpha$, $\beta$, and $\gamma$ are hyperparameters.  \cite{preload1, preload2, apl, quty} jointly optimize ladder selection and download control to increase optimization room, but optimization goal remains the same. 

There have been sporadic attempts and discussions in industry in the past to improve $QoS\to Profit$. For example, ~\cite{ruyi} attempts to obtain user preferences for video quality and fluency by asking each user to rate playback experience in a laboratory environment, and then set preference weight for users. ~\cite{sensei} designs a set of labeling methods to obtain user preferences for quality and fluency of videos and their segments through labeling on a crowdsourcing platform, so as to improve $QoE$ of videos to meet user preferences. 

Differences in our work: 

\begin{enumerate} [leftmargin=*]
    \item Experience optimization is no longer targeted at a single video, but rather at whole session. Performance metrics are expanded to $QoP$ which can directly points to $Profit$. Subordinate $QoP$ indicators will be established specificly to this task, which is decomposed into multiple aspects. For example, rebuffer will be decomposed into first frame rebuffer, rebuffer at beginning, middle or later, etc. 
    \item  Control actions are more sophisticated. For example, ladder selection requires distinct timing and goal. Download action requires not only choosing "what to download" and "how much to download", but also "where and how", and "what next after downloading". On this basis, joint decisions are made for available actions. Apart from joint decision of ladder selection and download, it includes either the two with other actions such as decoding and post-processing to maximize $QoP$. If we pick up specific algorithm based on input features, such as a model or decision tree, called decision function $Decider$, there are several cases of joint strategy described below: 
    \begin{enumerate}
        \item For decision of a certain control action, results in previous timeline can be input into $Decider$ as features, and predictive decisions can also be made in later timeline. 
        \item For a certain timing, if there are multiple actions, a joint $Decider$ can be established to output optimal result. 
        \item Regardless of timing, $Decider$ can be searched (trained) offline to obtain better method. 
    \end{enumerate}
    \item Filtering features based on common analysis method, a large number of personalized features need to be added in addition to common status like network and buffer, include (but not limited to) the following: 
    \begin{enumerate}
        \item Prediction of user item playtime, ...
        \item User and video first frame sensitive portrait, video quality sensitive portrait, video quality switching sensitive portrait, rebuffer sensitive portrait, power consumption sensitive portrait, traffic sensitive portrait, ...
        \item User current nearline and real-time video quality and rebuffer, temperature, volume, screen brightness, cpu, memory occupancy, traffic, ...
    \end{enumerate}
\item Finally, based on above features, $QoP\to Profit$ mapping can be updated synchronously to construct the sensitivity and preference $QoP\_sens^{it}_{u}$ for users and their separated dimensional experiences when watching items. And iterate corresponding decision functions. 
\end{enumerate}

In abstract terms, optimizing algorithms in whole streaming \& playback area is similar as above process, which can be divided into online (Algorithm~\ref{alg:Framwork}) and offline process of updating features, $QoP\_sens^{it}_{u}$, $Decider$ (Algorithm~\ref{alg:Framwork2}).

\begin{algorithm} [htb]
    \caption{Streaming \& Playback Decision}
    \label{alg:Framwork}
    \begin{algorithmic} [1] 
    \begin{footnotesize}
    \REQUIRE ~~\\ 
        User state:\  $U$; \\
        Item state:\ $Itl$;\\
        Context:\  $Ctx$;\\  
        UserAction:\  $UAction$;  \\
        Action List:\  $ACT$;   \\
    \ENSURE ~~\\ 
         \WHILE {user in App}
                \STATE $acts$ = $get\_actions (UAction)$  
                \STATE $U, Itl, Ctx$ = $update\_State ()$   
                \STATE 
                $QoP\_sens^{it}_{u}$ = $get\_QoP\_sens ()$ 
                \STATE  
                $ \begin{aligned}[t] acts\_params &  =  \mathrm{Decider_{acts}} (ACT, UAction, \\
                & QoP\_sens^{it}_{u}, Ctx,  U, Itl)
                \end{aligned}
                $ 
                \STATE $acts\_results$ = \textbf{Execute} ($acts, acts\_params$)  
                \STATE $ACT += acts, acts\_params, acts\_results$
            \ENDWHILE
    \end{footnotesize}
    \end{algorithmic}
    \end{algorithm}

\begin{algorithm} [htb]
        \caption{Decider Optimization}
        \label{alg:Framwork2}
        \begin{algorithmic} [1] 
        \begin{footnotesize}
        \REQUIRE ~~\\ 
            User state:\  $U$; \\
            Item state:\ $Itl$;\\
            Context:\  $Ctx$;\\  
            Training Data: $Data=\{x_i, y_i\}_{i=0}^N$; \\
            Collection of Deciders: \ $DC$; \\
        
        \ENSURE ~~\\

            \WHILE{True}
            
                \STATE  $Ctx, U, Itl$ = $extend\_State ()$  
                 \STATE $
                 \begin{aligned}[t] QoP&\_sens^{it}_{u}  =  update\_QoP\_sens (Ctx, U, Itl)
                 \end{aligned}
                 $
                 \STATE Deciders = $Heuristic\_Search (DC)$
        
                 \FOR {$\rm{Decider_k} \  \in$ \ Deciders} 
                    \STATE  $\rm{Decider_k}$ = $Optimize$ ($\rm{Decider_k}$)  
                \ENDFOR

            \ENDWHILE
            \end{footnotesize}
            \end{algorithmic}
            \renewcommand{\algorithmicrequire}{ \textbf{Decider Optimize Approach 1:}} 
            \begin{algorithmic} [1]
            \begin{footnotesize}
        \REQUIRE ~~\\
            
            \FOR{$b \in$ batch ($Data$)} 
                    \STATE  $\mathcal{L} (\hat \theta_k) = \frac{1}{b}\sum_i^b loss (y_i ,\rm{Decider_k} (\hat \theta, x_i, QoP\_sens^{it}_{u}))$
                    \STATE $\hat \theta_k = \hat \theta_k - \nabla \mathcal{L} (\hat \theta_k)$
                \ENDFOR
        \end{footnotesize}
         \end{algorithmic}
            \renewcommand{\algorithmicrequire}{ \textbf{Decider Optimize Approach 2:}} 
            \begin{algorithmic} [1]
            \begin{footnotesize}
        \REQUIRE ~~\\
            \FOR {$episode \in Data$} 
                    \STATE $s$ =  ($Ctx, U, Itl$), $a$ = $acts\_params$
                    \STATE $Q (s,a)$ = $\rm{Decider_k}$
                    \STATE $r = EstProfit (QoP, QoP\_sens^{it}_{u})$
                    
                    \FOR {each  step of $episode$} 
                        \STATE Choose a from s using policy derived from Q (s,a)   (e.g. $\epsilon-greedy$) 
                        \STATE  Take actions with a, observe $r$, $s{'}$
                        \STATE  $
                        \begin{aligned}
                        Q (s, a)  \leftarrow   Q (s,a)   + \alpha  [r + \gamma \max\limits_{a{'}}Q (s{'} ,  a{'}) - Q (s, a)]
                        \end{aligned}$
                        \STATE $s \leftarrow s{'}$
                    \ENDFOR
                \ENDFOR
        \end{footnotesize}
        \end{algorithmic}
        \end{algorithm} 
        
\noindent \textbf{Architecture}

Traditionally, design concept of multimedia frameworks such as DirectShow, Gstreamer, and FFMpeg is to modularize various components, use unified interfaces, flexibly build DAG like pipelines, improve reusability and consistency. 

\begin{figure}
    \centering
    \includegraphics [width=0.8\linewidth]{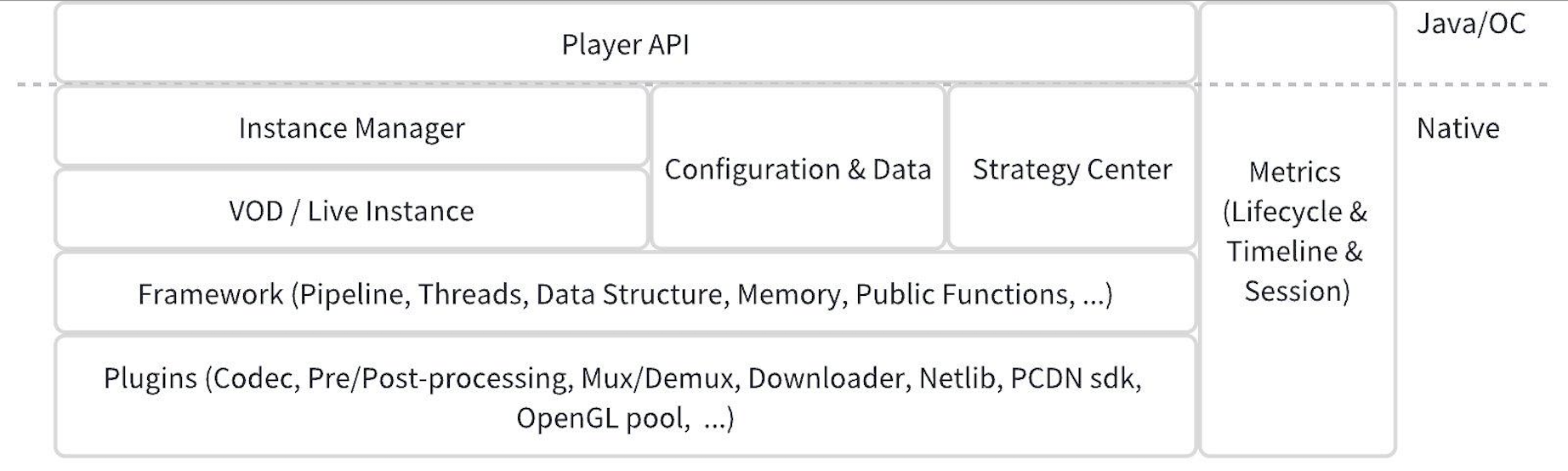}
    \caption{Player Architecture Schematic Diagram}
    \label{fig:api}
\end{figure}

However, regarding issues related to personalized streaming \& playback defined, since the top priority is to maximize user experience, starting point is different from that of players based on traditional multimedia frameworks. Therefore, we have established following player architecture, see Fig.~\ref{fig:api}.

\begin{itemize} [leftmargin=*]
 \item \textbf{Establish strategy center component for coordination}
 
 Strategy center is to control playback and download, including real-time and background calculation. The main purpose is that algorithmization will be long-term and frequent occurrence. Decision-making part of each component (i.e., $A_{top}$ in the previous paragraph) is moved to it and optimized in an coordinate manner, look at formula \eqref{equ:maxProfit2}: 
 
     \begin{footnotesize}
        \begin{equation}
            \begin{aligned}
            & \max_{\pi_{u} (A_{client}|S_{client})}Profit^{itl}_{u}\left  ( QoP^{itl}_{u}\left  ( \pi_{u},s_{u} \right )| s_{u} \right ) \\
            & >
            \max_{\pi_{u}^i (A_{client}^i|S_{client}^i)}^{}Profit^{itl}_{u}\left  ( QoP^{itl}_{u}\left  ( \pi_{u}^i,s_{u} \right )| s_{u} \right ) \label{equ:maxProfit2}
            \end{aligned}
        \end{equation}
    \end{footnotesize}
    \item \textbf{Establish globally visible configuration and data module}
    
        Configuration and data module are responsible for providing globally visible configuration and data access capabilities, which simplifies interactions between different components. Data mentioned here should include operating status data and intermediate calculation results within each module, representing state space. 
\item \textbf{Establish unified abstract framework}

        Unify design of each component and use same set of capabilities such as thread pool, lock, message, memory buffer pool, etc. It is similar to traditional multimedia framework. Only difference is to provide much more implementations for basic capabilities as choices
\item \textbf{Provide extreme adjustable capabilities aligned with $QoP$}

        Components in traditional multimedia frameworks also provide options to bring a basic level of flexibility. For example, libde265dec component of Gstreamer only allows maximum number of threads to change, which is far from enough. Users require "optimal" performance in their situations.
        Furthermore, all components should provide improved (and convertible) versions for major secondary performance indicators, or more refined configuration options for their local processes, for a component that provides demux capability, there should have high/medium/low memory usage, high/medium/low cpu usage versions, etc. Key is to be open and adjustable, meaning that algorithms in strategy center can dynamically adjust them. From optimization perspective, this can be considered to be improving $a_{ (i,k),j}$ and its corresponding $QoP_{ (i,k),j}$, expanding action space.
\end{itemize}

Overall speaking, our design increases complexity of program, but in exchange for higher performance improvement room and strategy coordination space to provide more personalized streaming and playback experience. 

\subsection{Personalized Quality Scheduling}

Usually, a video platform will use several CDN vendors for content distribution and disaster recovery, to consider how to better utilize the resource to deliver content with higher experience and lower cost. 

Previous approach is to make random allocations based on cost constraints and load balancing, or to make dynamic scheduling based on $QoS$. However, they didn't take into account user features, such as network preference of users, neither video features like popularity. 

Our approach generally includes defining priority of each request for each user based on playback status and preference for $QoP$, performing personalized quality scheduling to match best network resource under constraints. Below are formula for objective under peak and traffic mods.
\begin{itemize} [leftmargin=*]
    \item \textbf{Objective}
    \begin{itemize} [leftmargin=*]
        \item 95 peak billing mode, $Cost_{c}$ is \eqref{equ:95peak},    
        \begin{equation} 
    \begin{footnotesize}
    \begin{aligned}
Cost_{c} = \arg \min \max_{t_{95}}\{{ \sum_{r_i, c_j} B_{edge} (r_i, c_j) }  +  { \sum_{r_i, c_j} B_{bts} (r_i, c_j) }\}\label{equ:95peak}
\end{aligned}
\end{footnotesize}
 \end{equation}
 
\item Traffic billing mode, $Cost_{c}$ is  \eqref{equ:agg}, 
\begin{equation} 
\begin{footnotesize}
    \begin{aligned}
Cost_{c} = \arg \min \sum_{t}\{{ \sum_{r_i, c_j} B_{edge} (r_i, c_j) } + { \sum_{r_i, c_j} B_{bts} (r_i, c_j) }\} \label{equ:agg} 
\end{aligned} 
\end{footnotesize}
 \end{equation} 
 \item Object is \eqref{equ:cdn_schedule_object},
 \begin{equation} 
 \begin{footnotesize}
    \begin{aligned} \label{equ:cdn_schedule_object}
    & max (\sum_{r_i, c_j} LTV_{u} (QoP (r_i,c_j|p_j,b_j))  - Cost_{c}) \\
        & s.t.
        B (\pi_c (s (r_i))=c_j) = B_{all} * b_j,
        \forall c_j \in C  
    \end{aligned}
\end{footnotesize}
\end{equation} 
 \end{itemize}
\noindent Here, $C$ represents a collection of CDN vendor, $c_j$ is one of the vendors. $B$ represents bandwidth, while $B_{edge}$ means edge bandwidth and $B_{bts}$back-to-source bandwidth, $B_{all}$ covered two of them. $r_i$ represents specific request. $p_j$and $b_j$ represent unit price and bandwidth ratio for selected vendor. 
    
\end{itemize}

\begin{figure}
    \centering
    \includegraphics [width=0.8\linewidth]{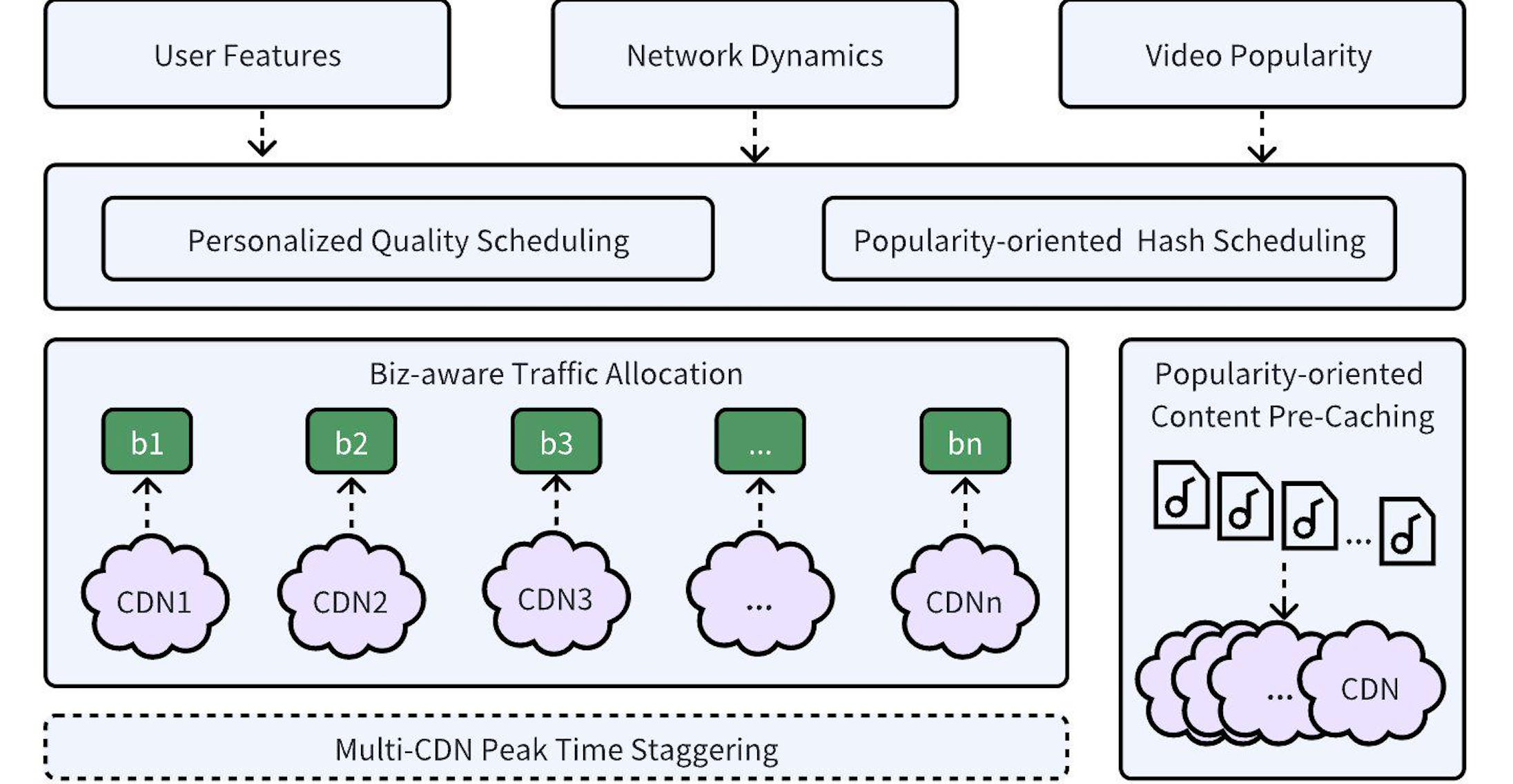}
    \caption{Personalized Quality Scheduling}
    \label{fig:scheduling}
\end{figure}

 By disassembling components of above objective function, each module can be iterated without affecting others. Personalized quality scheduling under cost constraint improves $LTV_{u}$, staggering peak height affects final billing cost by allocating $B_{edge} (r_i, c_j) $, video pre-cache reduces the back-to-source bandwidth $B_{bts} (r_i, c_j)$ by improving edge hit rate, as well joint optimization that considers both experience and cost, the overall picture shown in Fig.~\ref{fig:scheduling}

\noindent \textbf{Personalized Quality Scheduling} \label{subsec:scheduling}

The main challenge that CDN quality scheduling needs to solve is to predict service quality of resources in a dynamic network and make the best allocation based on user needs while meeting cost constraints. ~\cite{jiang2016cfa} proposed a resource quality prediction and decision-making method based on key features, but this method did not consider cost constraints and actual differences in user download needs. In 2022, we have proposed a quality scheduling method that considered cost and capacity constraints~\cite{DIG}. 

A download request $r_i$, in order to meet minimum requirement of network quality for playback fluency, is assumed to depend on user, video and playback status, extending previous definition to status of each request $s (r_i)$. For a given download request, service quality of vendor $c_j$ is defined as $QoP (c_j,s (r_i))$. Cost constraint is reflected in target magnitude of bandwidth. Assuming that total bandwidth is $B_{all}$, allocation to vendor $c_j$ should meet expected component ratio $B (c_j) = B_{all} * b_j$, where $b_j$ is bandwidth ratio of vendor $c_j$. Then the problem is to provide resources that meet quality requirements for each download request, that is, to design a scheduling strategy to maximize user experience, see \eqref{equ:maxQoP}. 
\begin{equation}  \label{equ:maxQoP}
\begin{footnotesize}
\begin{aligned}
& \arg \max_{\pi_c:s\to c} \sum_{r_i} LTV_{u} (QoP (\pi_c,s (r_i))|s (r_i)) \\
& s.t.
 B (\pi_c (s (r_i))=c_j) = B_{all} * b_j,
 \forall c_j \in C
\end{aligned}
\end{footnotesize}
\end{equation} 

Here, $LTV_{u}$ represents value brought by actual user experience after resource allocation. Note that traffic share is assumed to be fixed. Therefore, above strategy needs to meet vendor constraints to avoid excessive or insufficient resource allocation due to vendor quality. 

The definition of $s (r_i)$ needs to be determined in combination with user's sensitivity to playback performance and current status, refer to formula(\ref{equ:PsPObjective}). For scheduling algorithm, it is to predict CDN quality $QoP (c_j, s (r_i))$. A feasible way described in ~\cite{DIG}  is to adapt to network performance based on method of key feature grouping, real-time prediction and scheduling. 

\noindent \textbf{Biz-aware Traffic Allocation}

Usually, resource allocation ratio $b_i$ of vendor $i$ only considers service quality (e.g., availability, speed) and unit cost. For example, previous work \cite{buyya2008CDN, qosAwareCDN} mainly studies allocation method of user requests and internal node costs (bandwidth, storage) from perspective of CDN vendors, but lacks measurement of impact on real user experience. This section will discuss allocation. 

To illustrate the problem, define symbols first, CDN vendor $c_i$, unit price $p_i$, bandwidth traffic ratio $b_i$, $i \in \{1,2, ..., n\} $, $n$ represents number of vendors. Based on price and allocation, we can calculate cost, see  \eqref{equ:costcdn}, 
\begin{equation}  \label{equ:costcdn}
\begin{footnotesize}
COST_{bandwidth}=\sum p_i\times b_i \times B_{all} = B_{all} \times \textbf{p} \times \textbf{b}
\end{footnotesize}
\end{equation}
\textbf{p}, \textbf{b} are related to actual business negotiations. For example, increasing quantity level may result in price discounts. There are two factors, 
\begin{itemize} [leftmargin=*]
    \item \textbf{Experience value definition}: when scheduling strategy $\pi$ remains unchanged, relationship between experience value and resources is expressed as $LTV (QoP (\textbf{b}|\textbf{p})) $ (we believe quality is related to unit price and magnitude, that is, unit price and magnitude are main factors affecting experience). 
    \item \textbf{Disaster recovery constraints}: it is necessary to ensure that number of vendors is not less than $\eta$, considering that bandwidth capacity is limited, only under disaster recovery requirements are met, component objective is to optimize utility function (experience value-cost), like \eqref{equ:utility}, 
\begin{equation}  \label{equ:utility}
\begin{footnotesize}
\max_{\pi_{b}: \textbf{b}}\{LTV (QoP (\textbf{b}|\textbf{p}))  - COST_{bandwidth}\}
\end{footnotesize}
\end{equation}
\end{itemize}

Based on current allocation, AB experiments can verify utility functions under all methods, with idea of gradient descent, our allocation targets can be adjusted according to $max (\Delta Profit)$. However, unlike model training, experimental verification often takes several weeks, so it is impossible to traverse all parameters. It requires introduction of prior knowledge to assist decision-making, such as measuring the download speed and playback performance for each vendor, thereby reducing search space. 

\noindent \textbf{Popularity-oriented Hash Scheduling}

In addition to edge bandwidth, another part of cost comes from back-to-source bandwidth. In CDN architecture, edge hit refers to whether file requested is cached. If not, edge node needs to pull file from source. Usually, it will increase cost and reduce experience. 

Since edge hit rate of cold files is worse, it is hoped that heat change information of video files can be used to perform targeted popularity scheduling, cold files are centrally scheduled to a limited number of vendors and nodes, thereby improving edge cache hit rate. 

Back-to-source bandwidth can be expressed as \eqref{equ:bts}, 
\begin{equation} 
\begin{footnotesize}
{ \sum_{r_i, c_j} B_{bts} (r_i, c_j) } \label{equ:bts}
\end{footnotesize}
 \end{equation} 
Here, $r_i,  c_j$ represent a download request and corresponding resource respectively. If edge node of $c_j$ caches the file, then $B_{bts} (r_i, c_j) = 0$. Otherwise, corresponding back-to-source bandwidth will be generated. 

Quality or cost based scheduling mentioned above in industry dispatch file requests to vendors randomly or uniformly, which will further reduce file popularity from vendor perspective, resulting in a further decrease in edge hit rate. Therefore, to improve hit rate, the main solution is to distribute files only to limited vendors. However, if all videos are scheduled in targeted manner, risk is traffic can increase suddenly, so a reasonable solution is to predict file popularity (refer to chapter~\ref{subsec:VVP}), and divide files into hot and cold, such as selecting top N\% of cold files for quality scheduling. 

\noindent \textbf{Popularity-oriented Content Pre-Caching}

Common practice of vendors to deploy files is to cache files based on LRU like mechanism, in history, ~\cite{flexibleCDNCache} to allocate more requests to nodes with caches in traffic dispatching, ~\cite{caca} allocate traffic based on deployment strategy through historical popularity. To improve hit rate, a reasonable idea is to predict future popularity and deploy in advance. 

Composition of bandwidth cost can mainly be divided into download bandwidth (edge traffic) ${ \sum\limits_{r_i, c_j} B_{edge} (r_i, c_j) }$ and back-to-source (deployment) bandwidth ${ \sum\limits_{r_i, c_j} B_{bts} (r_i, c_j) }$. Objective settings of each are different for billing modes (such as bandwidth peak mode and traffic mode). 

Taking monthly 95-peak~\eqref{equ:95peak} billing mode as an example, if there are peaks and valleys in waveform throughout the day, we can use idle bandwidth resources in valley to deploy in advance, thereby improving hit rate especially in peak time. 

Reasonable solutions include: 
\begin{itemize} [leftmargin=*]
    \item Implement video ladder value calculation and further split regional operators for prediction (node services usually are not cross operators or regions, which will cause certain loss). 
    \item In order to avoid excessive node load caused by centralized deployment of hot files, use knowledge graph technology to characterize similarity of files, and pre-cache files according to the principle of maximum overall file similarity. 
\end{itemize}

\noindent \textbf{Multi-CDN Peak Time Staggering}

For CDN suppliers, peak billing mode is usually cost-effective than traffic mode~\cite{buyya2008CDN}. In case of having multiple CDNs, cost is the sum of 95 peaks. Thus, for video platforms, further cost optimization can be achieved by allocating user download requests and controlling bandwidth waveform between CDN vendors. Given total bandwidth $T$, billing of each vendor is reduced, Define bandwidth scheduling reuse rate $SRR$ as \eqref{equ:SRR}: 
\begin{equation} \label{equ:SRR}
\begin{footnotesize}
SRR = \frac{T_{95} - \sum t_{95} (c_i)} {T_{95}}
\end{footnotesize}
\end{equation} 

The higher $SRR$, more bandwidth saved. Generally, we believe that $SRR \geq 0$. When actual ratio of each CDN $c_i$ and the expected allocation $t_i$ account for same proportion of all vendors (no waveform intervention performed), $SRR = 0$. 

Controlling bandwidth by allocating ratios considering resource capacity limits of vendors, $t_i \leq T_i$. Stacking needs to make daily granularity decisions, based on overall waveform $\Lambda_T$ of each day, and according to stacking height and remaining free time of vendor (for example, 5\% of free time under 95 peak model), it is decided in which time interval to stack which vendors $C_{st} = \{c_1, ..., c_k\}$ and traffic ratio of the stack $T_{st} = {t_1,...,t_k}$, so as to maximize overall reuse rate, see  \eqref{equ:SRR2}, 
\begin{equation} 
\begin{footnotesize}
\begin{aligned}
& \arg\max_{C_{st},T_{st}}SRR \label{equ:SRR2} \\ 
& s.t. \sum t \sim \Lambda_T  \ \ \ t_i \leq T_i 
\end{aligned}
\end{footnotesize}
 \end{equation} 

\begin{figure}
    \centering
    \includegraphics [width=1.0\linewidth]{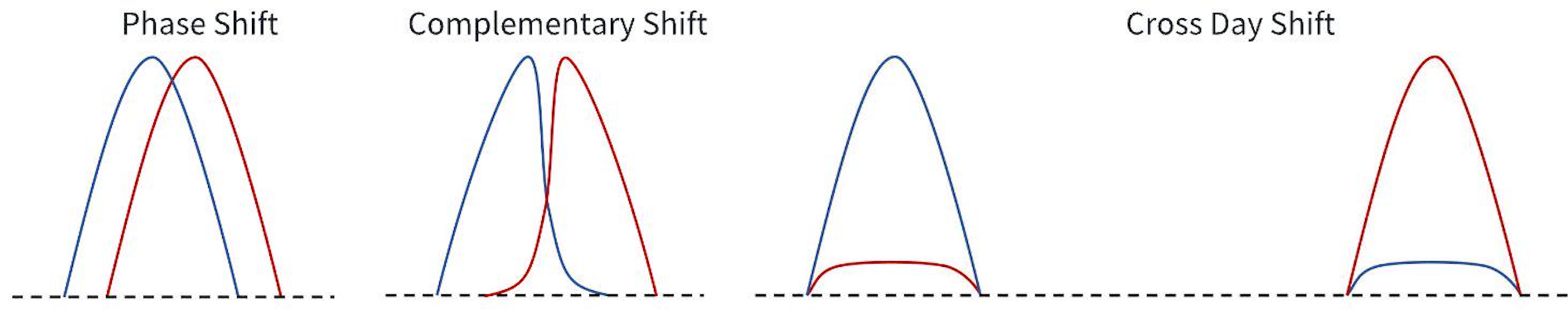}
    \caption{Schematic Diagram of CDN Peak-Shifting Stacking Method.}
    \label{fig:CDNPeakShifting}
\end{figure}

For time period, according to billing mode, predict time interval within 95 peak of monthly granular bandwidth. There are three ways to stack like Fig.~\ref{fig:CDNPeakShifting}. Considering complexity of implementation and minimizing the bandwidth jitter of CDN node in a single day, cross day shift is a better choice . 

\subsection{On Demand Delivery}

\noindent \textbf{On Demand  Service}\label{vods}

Sending more available video ladders to client can help to improve room of client, but at same time transmitting and data parsing has overhead. If useless ladder files sent, it has negative impact. So the question is how to select the best "possible combination" on delivery side, balance accuracy of both server and client, and consider cost of delivery action itself. 

Formal formula for this problem can be set up as \eqref{equ:delivery}: 
\begin{equation} 
\begin{footnotesize}
\begin{aligned} \label{equ:delivery}
\arg \min_{ \pi_d (A_{d} |s) }  \{ & \sum_{i=0}^{i<K} P^s_i * \min_{0\le j<K}  [replace\_cost^s_i  (j) *  (1-d_j)]\\
& +\sum_{i=0}^{i<K} d_i * deliver\_cost^s_i  \}
\end{aligned}
\end{footnotesize}
\end{equation}
\begin{itemize} [leftmargin=*]
\item  $K$ indicates a total of $K$ ladders. 
\item $A_{d} =\left \{ d_0,d_1,...,d_{K-1}\right \}$ indicates action space of delivery module, where $d_i\in \left \{ 0,1 \right \}$, $d_i=0$ indicates that ladder $L_i$ is not sent, and $d_i=1$ indicates to deliver the ladder. 

\item $s$ indicates ladder relative state of client predicted, as described in chapter \ref{subsec:PsP}. 
\item $P^s_i$ indicates probability that ladder $i$ is most suitable among all $K$ ladders in state $s$. The concept of "most suitable" implies difference between server-side and client-side strategies. Server side can theoretically use more global information, more complex and accurate algorithms, while client side can use device specific information and make decisions real-time. 
\item $replace\_cost^s_i (j)$ indicates profit and cost losses caused by not selecting ladder $i$ but selecting $j$ in state $s$ and most preferred is $i$. 
\item $deliver\_cost^s_i$ indicates transmission and other related overheads caused by sending ladder $i$ in $s$. 
\end{itemize}

In real world, since $K$ is generally not very large, for example, $K<100$, it is relatively easy to traverse $d_i$ after appropriate pruning. The main work is to estimate $P^s_i$, $replace\_cost^s_i (j)$and $deliver\_cost^s_i$ under state $s$. 
\begin{itemize} [leftmargin=*]
\item \textbf{Estimation of optimal ladder probability $P^s_i$}
\begin{itemize} [leftmargin=*]
\item $P^s_i$ represents probability that ladder $i$ is most suitable among all $K$ ladders in state $s$,  
\begin{equation}
\begin{footnotesize}
\begin{aligned}
P^s_i &=\prod_{j=0}^{j<K}P  [Profit^{it}_{u} ( QoP^{it}_{u} ( L_i,s_{u} )| s_{u} ) \\
& \ge Profit^{it}_{u} ( QoP^{it}_{u}  ( L_j,s_{u})| s_{u} ) ]  \nonumber
\end{aligned}
\end{footnotesize}
\end{equation}

\item There are two ways to estimate $P^s_i$: \\
\textbf{Inductive method:} Send all ladders to the client, obtain a set of state \& final ladder selection data pairs $<s,L_i>$, and then construct the $<s,P_i^s>$ data set through clustering and bucketing, and realize prediction of $P_i^s$ based on $s$. \\
\textbf{Deductive method:} Model $Profit$ directly based on information of state $s$ and ladder $L_i$, and calculate $P^s_i$. For specific modeling method, please refer to \ref{subsec:PsP}. 
\end{itemize}
\item \textbf{Estimation of ladder replace loss $replace\_cost^s_i (j)$} 
\begin{itemize} [leftmargin=*]
\item Mainly due to the difference between ladder $L_i$ and $L_j$, replacing $L_i$ with $L_j$ in state $s$ brings performance losses like fluency, aesthetics, load, and bandwidth cost. For specific method, please refer to modeling of $Profit$ and $QoP$ in \ref{subsec:PsP}. 
\end{itemize}

\item \textbf{Estimation of delivery loss $deliver\_cost^s_i$}
\begin{itemize} [leftmargin=*]
    \item Delivery loss $deliver\_cost^s_i$ mainly includes overhead caused by transmitting and parsing metadata of ladder $L_i$, which is related to metadata size carried by ladder $L_i$, as well as the network status and device performance. Therefore, $deliver\_cost^s_i$ can be simply expressed as 
    \begin{equation}
        \begin{aligned}
    &deliver\_cost^s_i= meta\_size (L_i )*\frac{1}{\alpha (dev)}*\frac{1}{\alpha (net)} \nonumber
        \end{aligned} 
        \end{equation}
        
    Here $\alpha (dev)$ and $\alpha (net)$ represent parameters that are positively correlated with device and network respectively. To simplify, they can be constants. 
\end{itemize}
\end{itemize}

~\\
\textbf{Context Service}\label{ctx_service}

Context service mainly predicts future trends for a certain indicator, for example, predicting bandwidth for next 5 minutes and percentile of bandwidth in whole day. Other indicators can include playback volume, network speed distribution, etc. The output is mainly used by other algorithms, to give corresponding time series information, which can be: 
\begin{itemize} [leftmargin=*]

    \item To influence scheduling and delivery results, for example, to reorder or filter delivered content. 
    \item To embed into delivered content for client. 
    \item To serve other backend model scenarios (see chapter~\ref{sec:UIAE}). 
\end{itemize}

The essence of context service is a kind of application of time series in multimedia area. Model of time series can be expressed as  
$y_{t:t+N}{'}=f (y_{t-k:t-1},x_{t-k:t-1},s)$. $y_i$ represents real target at the $i$ moment, $s$ represents static feature, such as CDN vendors, $x$ is time series feature, and $y'_{t:t+N}$ represents estimated value at the $N$th moment in the future at $t$. An initial method can consider moving average of static factors, or try complex TSA method. 
\section{User-Item Aware Encoding} \label{sec:UIAE}
\subsection{Problems \& Framework}

for a specific video (item) $i$, which versions should be transcoded to achieve global optimality.

Providing multiple bitrates for a video so that consumers can adaptively choose to play has become a standard paradigm today. However, a core issue has not been well addressed, for a specific video (item) $i$, which version should be transcoded to achieve global optimality. See formula \eqref{equ:L}:
\begin{equation} \label{equ:L}
\begin{footnotesize}
\mathcal{L}= \arg\max_{\mathcal{L}} \sum_{i \in I , L
 \in \mathcal{L}} \  \sum_{u \in U} Reward (i\rightarrow L | u)
\end{footnotesize}
\end{equation}
Where $i\rightarrow L$ indicates that video $i$ is transcoded into ladder group $L$, and $\mathcal{L}$ is the ideal global optimal ladder group. $U$ represents all users, $I$ represents all videos, $Reward$ is revenue brought by transcoding. 

Transcoding strategy perceives the context of each user, and combine personalized features to calculate ladder $L$ that best suits every user and transcode in real time. However, due to high cost and delay, huge search space, it is no possible to launch with this logic. 

Instead, we conducted the following approximations:
\begin{itemize} [leftmargin=*]
    \item \textbf{Focus on popular videos:} Short video applications often have top 1\% of videos accounting for about 70\%+ of playback volume, tail videos have short life cycle and low ROI. Based on popular videos $V_{top}$, volume can be reduced from billions to tens of millions. 
    \item \textbf{Transcode in advance:} We estimate distribution of consumption users $U_g$ in the future, calculate optimal $L$ on $U_g$,. It is necessary to continuously update $L$ to adapt to consumption distribution shifting. 
    \item \textbf{Restriction and throughput based:} As thinking of resource and quota, we'd like to set up task queues adopts the weighted-batch mode. 
\end{itemize}

Base on above approximation, update \eqref{equ:L} to \eqref{equ:L2}:
\begin{equation}
\begin{footnotesize}
\begin{aligned} \label{equ:L2}
\mathcal{L} &\approx \arg\max_{\mathcal{L}} \sum_{\substack{i \in I_{top} \\  L \in \mathcal{L}}}  Reward (i\rightarrow L|U_g)   \\
& s.t \sum_{i \rightarrow L} cpu (i \rightarrow L)<=Cores\\
\end{aligned}  
\end{footnotesize}
\end{equation}

$I_{top}$ represents top video set, The core idea is to optimize hot videos and estimate the distribution of future consumer groups $U_g$, and finally calculate the ladder group $L_g$ suitable for $U_g$ with limited cpu $Cores$. 

As mentioned, the $Reward$ for a video in this article is defined as follow formula  \eqref{equ:reward}: 
\begin{equation} \label{equ:reward}
\begin{footnotesize}
Reward = \Delta LTV (L) - \Delta COST (L) = Profit (L)
\end{footnotesize}
\end{equation}
$ \Delta LTV (L)$ is $LTV$ change of  $L$. $\Delta COST (L)$ is cost change caused by  $L$, including bandwidth, computing, storage, etc. If the difference regarded as $Profit$, combined with consumer, $Reward$ optimization goal should be like \eqref{equ:reward2}: 
\begin{equation}
\begin{footnotesize}
Reward = Profit (QoP|L;U_g)  \label{equ:reward2}
\end{footnotesize}
\end{equation}

\begin{figure}
    \centering
    \includegraphics [width=1.0\linewidth]{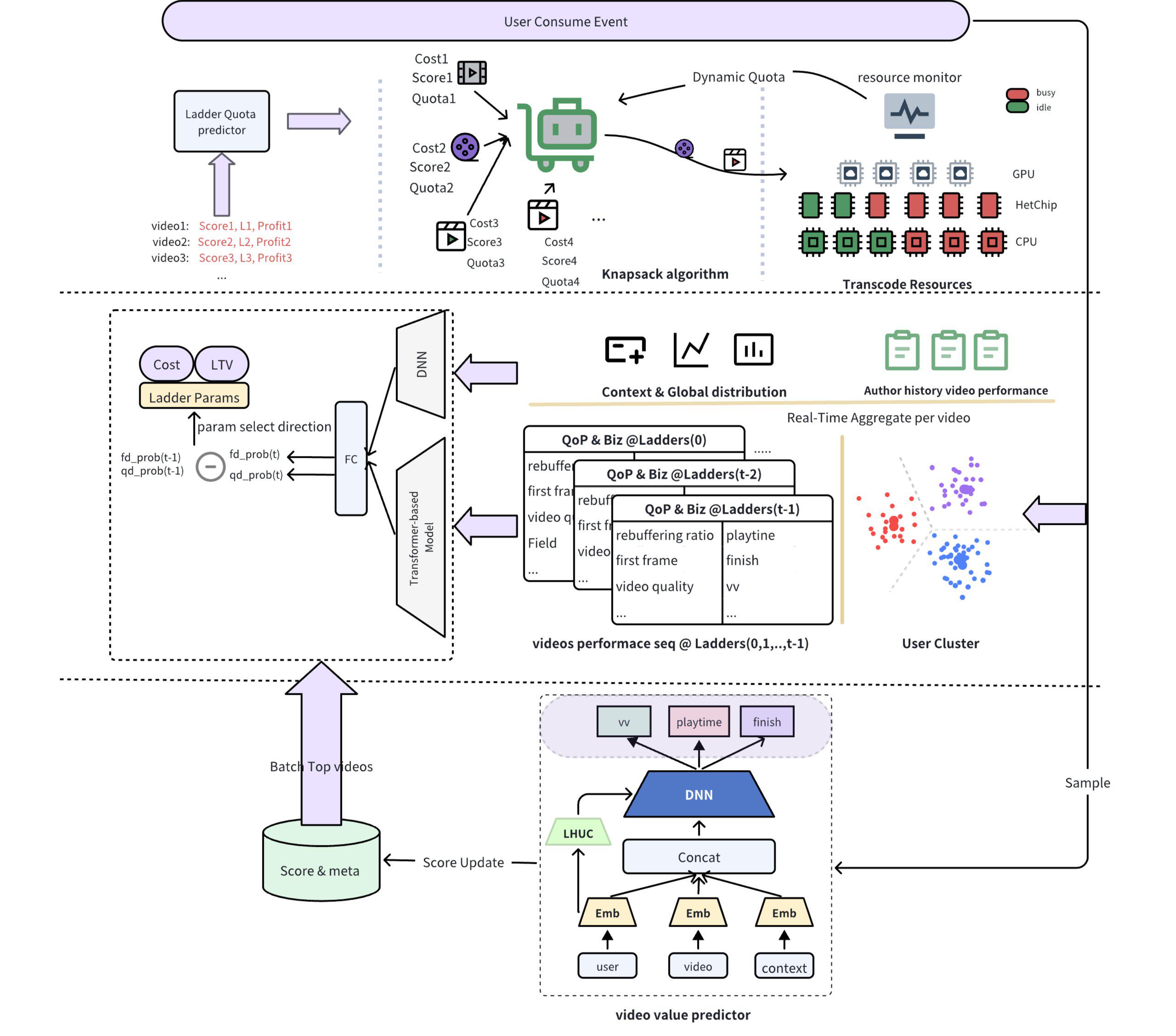}
    \caption{UIAE Detail Framework}
    \label{fig:UIAEDETAIL}
\end{figure}

We improve $Profit$  by optimizing $QoP$. The overall UIAE algorithm framework is shown in Fig.~\ref{fig:UIAEDETAIL}.

In the following chapter~\ref{subsec:VVP}, we introduce video value prediction model that is used in multiple scenarios. In chapter~\ref{subsec:UIAE}, introduce the UIAE for personalized ladder group. In chapter~\ref{subsec:RAM}, discuss how to maximize $Reward$ from a global view with limited resources. In chapter~\ref{subsec:Experiment_Methods}, talk about evaluation methods.

\subsection{Video Value Prediction} \label{subsec:VVP}

Value prediction plays an important role in decision-making and operation of business and society~\cite{influenza, inflationForecasts, deepar}. Meta has also published related work~\cite{Chess}, ~\cite{metaEncoding}, revealing how to select valuable videos. Value prediction modeling can be divided into two types. One is time series model. Backbone is mainly based on CNN \cite{TCN, borovykh2017condwavenet}, RNN/LSTM \cite{deepar, wang2019deep}, and Attention \cite{TFT, informer}, which is more suitable for periodicity scenarios. Another is to use deep networks such as DNN for direct regression modeling \cite{Chess, zhou2015video}, which is more versatile. 

The challenges faced in our value prediction are mainly:
\begin{itemize} [leftmargin=*]

    \item \textbf{Long tail distribution}: The proportion of high-value videos is extremely low, meanwhile, they account for the vast majority of playback volume. The regression model will be biased towards high-value videos, so the Loss function needs to be carefully designed. 
    
    \item \textbf{Short life cycle}: Short videos have short life cycle and diverse consumption trends, most follow geometric distribution. The consumption distributions of different categories of videos also vary widely.
\end{itemize}

The video value model has multi heads for various targets, e.g, view volume, play time, likes, downloads of multiple time windows. We uniformly use $\hat y$ to represent predicted target and $y$ to represent ground truth, see \eqref{equ:video_value_model}. 
\begin{equation}
\begin{footnotesize}
\hat y = f (item;ctx) \label{equ:video_value_model}  
\end{footnotesize}
\end{equation}
\begin{itemize} [leftmargin=*]

\item Partial feature examples: 
    \begin{itemize} [leftmargin=*]
        \item Author features: author ID, activity, number of posts, number of fans, number of views, etc.
        \item  Video features: music ID, content, duration, playback volume, like number, vv growth rate, etc.
        \item  Context features: time, holidays, etc.
    \end{itemize}
\item Various loss design:
    \begin{itemize} [leftmargin=*]
        \item MSE
  \begin{equation} 
  \begin{footnotesize}
Loss = \frac{1}{2}||y - \hat y||^2
\end{footnotesize}
\end{equation}
\item Huber Loss
\begin{equation}
\begin{footnotesize}
Loss_{\delta}=
    \left\{\begin{matrix}
        \frac{1}{2} (y - \hat{y})^{2} & if \left |  (y - \hat{y})  \right | < \delta\\
        \delta  (|y - \hat{y}| - \frac1 2 \delta) & otherwise
    \end{matrix}\right.
\end{footnotesize}
\end{equation}
\item Weighted log loss~\cite{covington2016deep}
  \begin{equation}
  \begin{footnotesize}
Loss=-ylogp - log (1-p)
\end{footnotesize}
\end{equation}

\end{itemize}
\item Model structure: one of video value model structures used online before, DNN+LHUC+BIAS, as video value predictor component shown in Fig.~\ref{fig:UIAEDETAIL}.

\item \text{Evaluation indicators}: REC-AUC~\cite{bi2003regression}, MAE~\cite{willmott2005advantages}. Table~\ref{table:Video_Value_Metrics} contains a comparison of this design with rule-based top vv strategy (based on view number) and BigV strategy (based on fans number). 
\end{itemize}

\begin{footnotesize}
\begin{table}
    \centering
    \caption{
    EVALUATION RESULTS
    } \label{table:Video_Value_Metrics}
    \begin{tabular}{p{2.3cm}p{1.0cm}p{1.5cm}p{0.7cm}}
    \hline
        \textbf{Loss} & \textbf{Model} & \textbf{RecAuc} & \textbf{MAE} \\
    \hline
    - & Topvv & 76.00\%  & -\\
    - & BigV & 60.00\% & - \\
    MSE & Ours & 89.50\%  & 0.507\\
    Huber & Ours & 89.70\% & 0.496 \\
    Weighted LogLoss & Ours & 89\% & 0.574\\
    \hline
    \end{tabular}

\end{table}
\end{footnotesize}
\subsection{User-Item Aware Encoding} \label{subsec:UIAE}
\begin{figure*}
    \centering
    \includegraphics [width=1.0\linewidth]{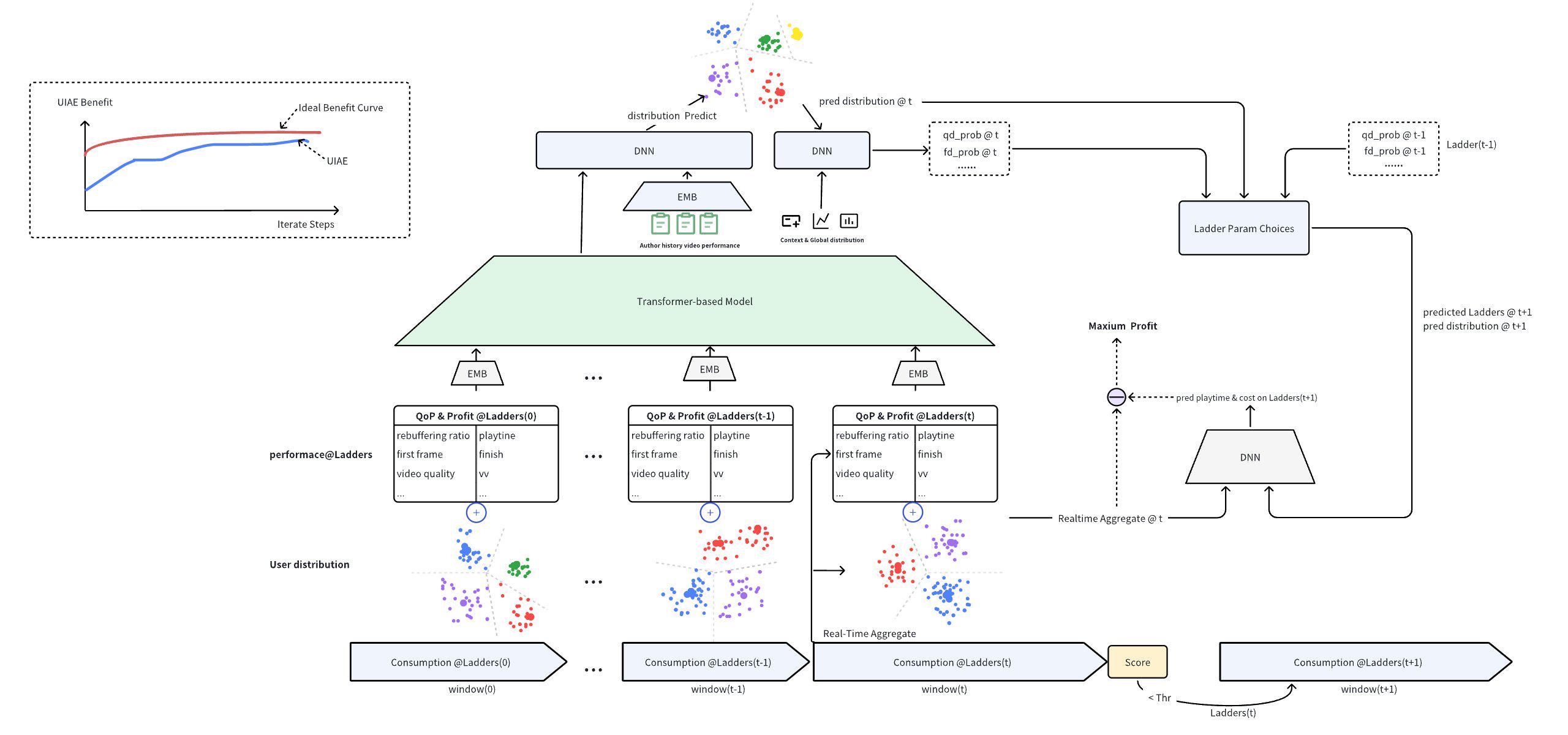}
    \caption{User-Item Aware Encoding Solution}
    \label{fig:UIAESolution}
\end{figure*}

It is difficult to get feedback about $L$ directly in playback scenario. A feasible approach is to divide the whole process into windows, and predict the optimal $L$ for next window with all past results, continuously update $L$ to ensure that $Reward$ is constantly approaching ideal value, see left top of Fig.~\ref{fig:UIAESolution}. The figure also describes the whole solution. Based on \eqref{equ:L2}, considering $QoP$ impact of transcoding, that is: 
\begin{equation}
\begin{footnotesize}
    \begin{aligned}
     \label{equ:reward3}
    Reward \! =\!Profit (QoP(r,d)|L;U_g)
\end{aligned}
\end{footnotesize}
\end{equation}
Here, $r$ is predicted bitrate, $d$ is video quality.

Set a consumption window ($vv_{window}=step$) and a video value threshold $score_{th}$. Execute following steps in each consumption window:
\begin{itemize} [leftmargin=*]
      \item Predict video value $score$, if $score$ is less than $score_{th}$, then withdrawn the current ladder calculation  (i.e., $Ladder (t)$ will still be used in $t+1$ stage). 
      \item Calculate $QoP (t)$ and $Profit (t)$ within window $t$ and consumer clustering, get user distribution $U_g$ (t) on window $t$. 
      \item Based on $input =  [ (U_g (0), QoP (0), Profit (0)); \\
       (U_g (1), QoP (1), Profit (1)); ...; (U_g (t), QoP (t), Profit (t))]$ to estimate $Ug (t+1)$, at same time, combined with input features such as $context$ and the consumption history of author works, estimated delta of $QoP$, like video quality preference $qd\_prob (t+1)$ and fluent preference $fd\_prob (t+1)$, etc., to guide the parameter selection of future ladders. 
      \item  Ladder parameter selection: 
      \begin{itemize} [leftmargin=*]
        \item According to $qd\_prob (t+1)$ and $fd\_prob (t+1)$ vs. $qd\_prob (t)$ and $fd\_prob (t)$, $L (t)$ and $U_g (t+1)$, calculate ladder group optimization direction (quality or fluency) among optional ladder parameters. 
        \item Calculate $Profit$ based on the estimated $Cost$ and $playtime (t+1)$ , which is based on potential ladder group $predLadder (t+1)$ produced by step 1. 
        \item According to equation \eqref{equ:reward2}, the largest ladder group of $Reward$ will be selected for $Ladder$. 
      \end{itemize}
    \end{itemize}

Encoding options can be found in supplementary materials. Given some options, we can predict video compress result (like common CAE method \cite{CAE-Netflix, mico2023per}), to support UIAE and Resource Allocation Model in chapter~\ref{subsec:RAM}. 

\subsection{Resource Allocation Model} \label{subsec:RAM}
When a video's $L$ is predicted, it should be transcoded and consumed. However, from global perspective, transcoding scheduling is not a all-in operation. 
\begin{itemize} [leftmargin=*]   
    \item Limited resource: It is impossible to meet all requests at same time, and transcoding cost of each video vary greatly. 
    \item Heterogeneous resource: Not only x86 CPUs, but also items like Arm CPUs, FPGAs, GPUs, and customized chips, each with its own performance and cost. 
    \item Extended impact: Other costs like CDN bandwidth should be considered for all $L$.
\end{itemize}

For transcoding scheduling, objective can be set to \eqref{equ:global_objective}: 

\begin{equation}  
\begin{footnotesize}
\begin{aligned} \label{equ:global_objective}
& \max \sum\limits_{\substack{i \in I_{top}   L \in \mathcal{L}}}  Reward (i\rightarrow L|U_g)  \\
& = \max \sum\limits_{\substack{i \in I_{top} \\ L \in \mathcal{L}}} \! \Delta LTV (QoP|L;U_g) \\ 
& -  \Delta Calc(L) - \Delta BW (L,U_g) - \Delta Store (L)  \\
& s.t \sum\limits_{i \in I_{top}} cpu (i \rightarrow L)<=Cores 
\end{aligned} 
\end{footnotesize}
\end{equation}

In formula, with Ladder group $L$, $\Delta BW$ is the change of bandwidth. $\Delta Store$ indicates the change of storage. $Cores$ indicates total number of available resources, generally a constant. $i\rightarrow L$ indicates that video $i$ has been transcoded to $L$ . $cpu (i\rightarrow L)$ represents CPU resources by video $i$ to  $L$.
\begin{itemize} [leftmargin=*]
    \item $BW$ indicates bandwidth consumption of group $L$, see \eqref{equ:BW}: 
\begin{equation}
\begin{footnotesize}
BW = Br (L, p (L|u))  \label{equ:BW}
\end{footnotesize}
\end{equation}
    $p$ indicates proportion of ladder $L$ consumption during peak reduction phase, and $Br$ is weighted estimated bandwidth based on ladder group. 

    \item $Calc (L)$ is used to estimate computational cost for given group of ladders, see  \eqref{equ:Calc}: 
\begin{equation}
\begin{footnotesize}
Calc (L) = Cr (L, process\_type, resource\_type) \label{equ:Calc}
\end{footnotesize}
\end{equation}
  $process\_type$ indicates processing method, including selection of encoder type, processing chain, parameters, etc., $resource\_type$ indicates type of resource used, such as CPU, FPGA, GPU, etc., and $C_r$ function is used to estimate computational cost of ladder group. 
    \item $Store (L)$ is used to estimate storage cost for given $L$, while $bitrate_l$ and $video\_dur_{l}$ represents ladder file bitrate and duration, see \eqref{equ:Store}: 
\begin{equation} \label{equ:Store}
\begin{footnotesize}
Store (L) = \sum_{l \in L} bitrate_{l} * video\_dur_{l}
\end{footnotesize}
\end{equation}
\end{itemize}

Look closely at the objective, it is an optimization problem with constraints (backpack problem), but it will fall into following problems when trying to solve it directly:
\begin{itemize} [leftmargin=*]
\item  It is difficult to accurately quantify available resources. Scheduling, load balancing, and buffer strategies among data centers will affect estimation of available resources. 
\item There are tens of millions of popular videos, the complexity is too high. 
\end{itemize}

This allows to use near-stream processing, a batch is processed at a time and available resources redefined. See  \eqref{equ:global_objective_nearstream}: 
\begin{equation}
\begin{footnotesize}
\begin{aligned} \label{equ:global_objective_nearstream}
 s.t  \sum\limits_{i \rightarrow l} Quota (i \rightarrow L)<=TopK_b 
\end{aligned}
\end{footnotesize}
\end{equation}
$TopK_b$ indicates current transcoding quota that can be used, defined as transcoding cost required to specify parameter level for 1 second of transcoding. For encoding time estimation of a new video file, features mainly include reference video, reference ladder parameters, pre-coding results of new video, new parameters, and the regression model can predict quota required $L$. $TopK_b$ can be dynamically adjusted by PID~\cite{PID} or other cybernetic algorithms.

\subsection{Experiment Methods} \label{subsec:Experiment_Methods}
Beyond common AB test, we constructed additional experimental capabilities to support iteration of above algorithms.

First, due to resource constraints, transcoding experiments usually need to be performed on a small part of videos, so it is necessary to design labeling capabilities to establish a mapping relationship between user-level AB experiments and video strategies.
Some basic requirements are as follows: 
\begin{itemize} [leftmargin=*]   
    \item Label the strategy for each group's output files. 
    \item Outputs can be filtered by aligned user groups, according to transcode time. 
    \item Support strategies for distinguishing timing. 
    \item For requests, item tags can be returned according to experiment user group. 
    \item Support dynamic division of resource pools to ensure fairness. 
\end{itemize}

\begin{figure}
    \centering
    \includegraphics [width=0.75\linewidth]{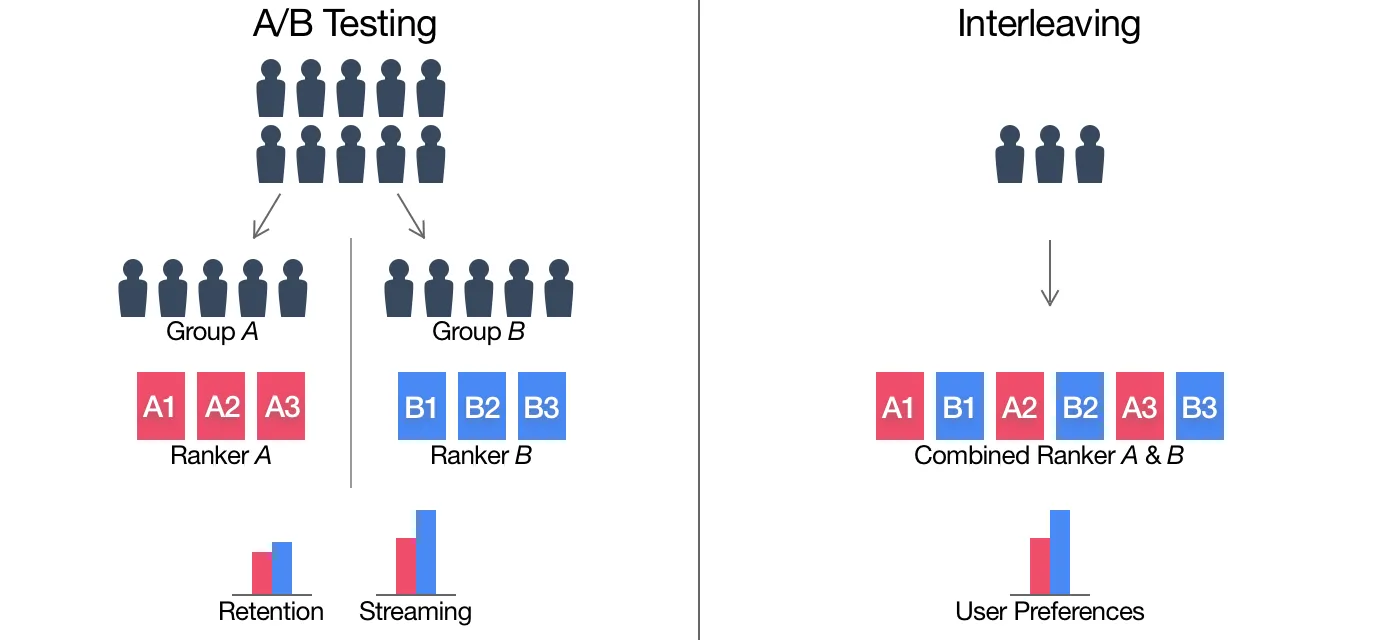}
    \caption{AB Testing and Interleaving~\cite{Interleaving}}
    \label{fig:Interleaving}
\end{figure}

Secondly, method of eliminating recommendation influence should be supported, impact to users should be distinguished in experimental. Interleaving experiment method in search scenarios~\cite{Interleaving} is often mentioned that mixing control and treatment to be compared and sending them to same user will bring better evaluation in some scenarios: 
Above Fig.~\ref{fig:Interleaving} shows general flow of interleaving experiment, we can alternately or randomly assign treatment and control strategies to the videos in a feed, the user sees both strategies.

Since video is played randomly, it can be approximated that this is a comparison of viewing effect of two strategies by same user. 

\begin{figure}
    \centering
    \includegraphics [width=0.9\linewidth]{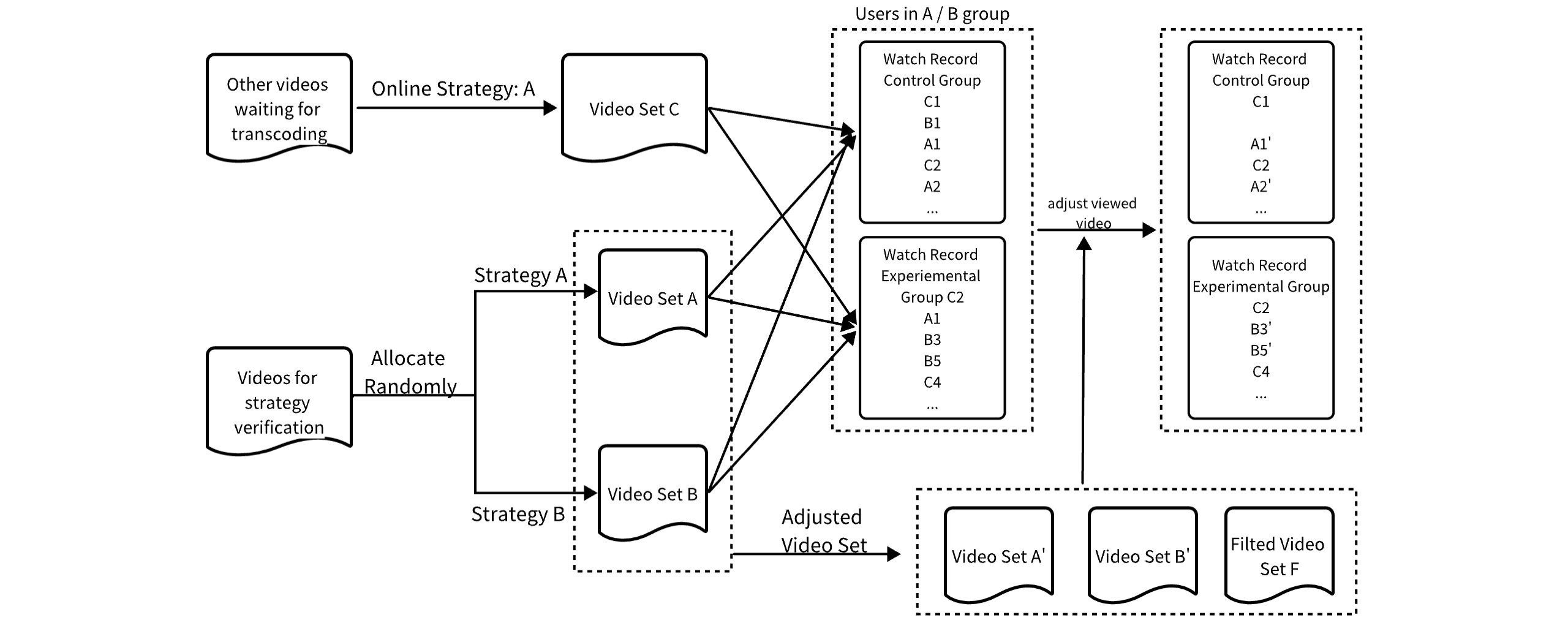}
    \caption{Quasi Experimental Solution}
    \label{fig:QuasiExperimental}
\end{figure}

Finally, when evaluating the effect of a transcoding strategy, above method will create different results for a source, then compare business effects of AB group, which requires large resources and long period. We designed a new method called "Quasi Experimental", which can quickly and accurately evaluate value of strategies while saving resources. The solution is to artificially construct experimental group, control group and corresponding playback behavior, business indicators, performance indicator. Essentially it's an application of causal inference. See details in 
Fig.~\ref{fig:QuasiExperimental}.

\begin{itemize}  [leftmargin=*]

\item \textbf{User dimension:} Users in experimental and control groups are randomly distributed as much as possible.
\item \textbf{Video dimension:} Select the video to be transcoded, randomly divide it into subsets A and B, and adjust the two video sets to meet the distribution of key dimensions, such as playback duration, video category viewed, video consumption resolution, etc. 
\item \textbf{Adjust viewing history:} Adjust playback record of the day, delete viewing records of users in group on videos in set A, and delete viewing records in control group in set B, so as to ensure that videos watched by users in control group are from strategy A (online strategy), and some watched by users from strategy B (to compare).

\item \textbf{Benefit estimation:} In experimental group $T$ and control group $C$, let the statistical value of observed metrics (e.g., playtime) on video set $X$ be expressed as $T_X$ and $C_X$, then we have formula \eqref{equ:Delta}: 
\begin{equation} \label{equ:Delta}
\begin{footnotesize}
\begin{aligned}
\Delta & =T_C-C_C+ (T_{B'}*\frac{||A+B||}{||B'||}-C_{A'}*\frac{||A+B||}{||A'||}) \\ 
& \approx T_C-C_C +  (T_{B'}-C_{A'})*\lambda  
\end{aligned}
\end{footnotesize}
\end{equation}
Here $\lambda$ is constant used in simplified calculation, usually between 2 to 3. 
\item \textbf{Performance estimation:} Based on results of metric adjustments we can get an estimation, see \eqref{equ:Delta2}, 
\begin{equation} \label{equ:Delta2}
\Delta=T_{C+B'}-C_{C+A'}
\end{equation}
\end{itemize}

\section{Foresight Based Publishing} \label{sec:FP}
\subsection{Problems \& Framework}
\begin{figure}
    \centering
    \includegraphics [width=1.0\linewidth]{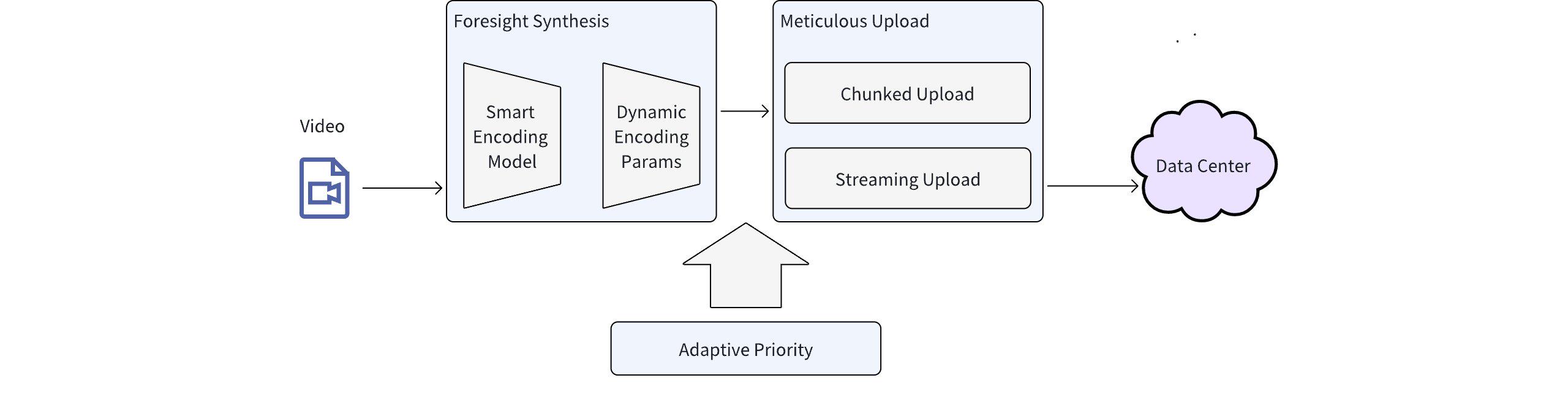}
    \caption{Publishing Optimization Process}
    \label{fig:publish_opt}
\end{figure}

Since material selection and editing process of publishing is too personal, we won't discuss it. Assuming that user has completed editing, this chapter will focus on common decision-making matters, which are divided into three parts based on synthetic encoding, uploading, and priority setting, as shown in Fig.~\ref{fig:publish_opt}.

Synthesis quality will affect transcoded results, and thus affect playback, publishing optimization should consider impact from both publishing and consumption. The goal is referenced as formula \eqref{equ:publish} : 
\begin{equation} \label{equ:publish}
\begin{footnotesize}
    \begin{aligned}
    \arg\max_{{\pi_{pub}
     (a_{pub}|s_{pub})}} & \sum_{u \in U_p}
    \overbrace{Profit (QoP ( \pi_{pub},s_{pub}))}^{publish}  \\ 
    & + \sum_{u \in U_c}  \overbrace{Profit (QoP ( \pi_{cons},s_{cons})}^{consume}
\end{aligned}
\end{footnotesize}
    \end{equation}

\begin{itemize} [leftmargin=*]   
    \item The first term in  \eqref{equ:publish} is publish experience of $U_p$. $\pi_{pub}$, $a_{pub}$, $s_{pub}$ represents strategy, action and state of publishing.
    \item The second term  in  \eqref{equ:publish} is consume experience of $U_c$. $\pi_{cons}$, $s_{cons}$ represents the strategy, state of consumption.
\end{itemize}

In our work, trade-offs for experience optimization objective can be divided into two categories: 
\begin{enumerate} [leftmargin=*]
\item There is clear impact or exchange between $Profit_{publish}$  and $Profit_{consume}$. It is necessary to make a forward-looking estimation of consumption. There are different approaches: 
\begin{enumerate} [leftmargin=*]
  \item Estimate consumption impact and merge it into publishing modeling, that is, estimate the personalized coefficient $\alpha_{u,i}$ based on publisher $u$ and submitted content $i$, and refer to this coefficient to correct the decision. It majorly impacts the scenario in chapter~\ref{subsec:Foresight_Synthesis}. 
  \item Directly estimate the two impacts. Due to the extremely long chain, it is difficult to accurately model, so it's only used when the consumption $QoP$ can be estimated, described in chapter~\ref{subsec:Adaptive_Priority}.
\end{enumerate} 
\item There are many tasks that only affect publishing. On top of that, the cost required for uploading is usually insignificant, such as scenario in \ref{subsec:Meticulous_Upload}. In this case, we can only optimize $LTV_{publish}$, that is, the ideal formula \eqref{equ:publish} degenerates into a local indicator that focuses on publishing scenario, and adds a constraint that consumption experience $LTV_{consume}$ cannot be negative. 

\begin{equation} \label{equ:publish_simple}
\begin{footnotesize}
    \begin{aligned}
\arg&\max_{{\pi_{pub}
 (a_{pub}|s_{pub})}}  \sum_{u \in U_p}
\Delta LTV_{publish} (QoP ( \pi_{pub},s_{pub})) \\ 
&s.t. \ \Delta LTV_{consume} \ge 0
\end{aligned}
\end{footnotesize}
\end{equation}
\end{enumerate}

There are multiple impact paths in publishing. Examples as follows: 
\begin{enumerate} [leftmargin=*]
\item Optimize video quality of the published content $ContentQuality$ to improve author satisfaction and get $LTV$ increase, like \eqref{equ:publish_idea}, \label{path1}
\begin{equation}
\begin{footnotesize}
\begin{aligned} \label{equ:publish_idea}
 \pi (a|s): \ & ContentQuality \uparrow \ \to LTV \uparrow  \\
\end{aligned}
\end{footnotesize}
\end{equation}

\item Optimize the publishing duration $PublishDuration$, thereby increasing success rate $PublishSuccess$, and thus achieving $LTV$ increase, that is \eqref{equ:publish_idea2}, \label{path2}
\begin{equation} \label{equ:publish_idea2}
\begin{footnotesize}
\begin{aligned}
 \pi (a|s):& PublishDuration \downarrow \ \to PublishSuccess \uparrow  \ \ \to LTV  \uparrow  \\
\end{aligned}
\end{footnotesize}
\end{equation}
\item Directly optimize publish success rate to get $LTV$ increase, that is \eqref{equ:publish_idea3},  \label{path3}
\begin{equation}
\begin{footnotesize}
\begin{aligned} \label{equ:publish_idea3} 
 \pi (a|s): \  &  PublishSuccess \uparrow  \ \ \to LTV  \uparrow \\
\end{aligned}
\end{footnotesize}
\end{equation}
\end{enumerate}

Chapter~\ref{subsec:Foresight_Synthesis} is mainly based path~\ref{path1} and ~\ref{path2}, chapter~\ref{subsec:Meticulous_Upload} based on path~\ref{path2} and chapter~\ref{subsec:Adaptive_Priority} based on path~\ref{path3}. 

Moreover, similar to chapter~\ref{subsec:PsP}, we construct personalized publishing preferences $QoP\_sens_{u, pub}$ , build features and portraits, then iterate optimization process. This chapter focuses on introducing the problems that can be personalized, objective trade-offs, decision matters and impact path, will not address details of solutions again. 
\subsection{Foresight Synthesis}  \label{subsec:Foresight_Synthesis}

When user finished creating, the material needs to be synthesized and encoded. Traditional approach is to encode the material based on pre-configured synthesis parameters, like filters, effects, etc. However, the pre-configured parameters are averaged that ignore each user's expectation, such as whether a quick upload or better video quality (which is beneficial for subsequent consumption). Network, device and context will also change user expectations. Based on it, we fully tap into user and item features, and make intelligent predictions on encoding methods.

In synthesis stage, we mainly follow path~\ref{path1} and \ref{path2} to maximize the benefits. A feasible idea is to decompose the process into two modules, which we call Smart Encoding Model and Dynamic Encoding Parameter, and optimize them one by one. Each module is independent to each other. The Smart Encoding Model is to select encoding mode, and Dynamic Encoding Parameter is to select parameters when encoding mode determined. Since the objectives of the two sub-problems are the same, we can also make a joint decision. 

\noindent \textbf{Smart Encoding Model}

For encoding mode selection problem, we define $a\in\!A_{enc}\!=\!\{soft\ enc, hard\ enc,skip\ enc\}$, the actions include software encoding, hardware encoding, and skip encoding. Since users are independent on this issue, it means $\arg\max \sum\Delta LTV\!=\!\sum \arg\max \Delta LTV$. Define the upload duration as $UploadDuration$ and encoding duration as $EncodingDuration$, the  objective is like formula \eqref{equ:publish_ltv}:
\begin{equation}
\begin{footnotesize}
\begin{aligned} \label{equ:publish_ltv}
& \arg  \max_{{\pi
 (a|s)}} 
  \Delta LTV (QoP ( \pi,s))\\   
& \approx \arg  \min_{{\pi
 (a|s)}} PublishDuration (\pi,s) \\ 
 & = \arg \min \limits_{{\pi(a|s)}} \max \{UploadDuration , EncodingDuration\}| (\pi,s)
\\  
& s.t. \ \Delta LTV_{consume} \ge 0
\end{aligned}
\end{footnotesize}
\end{equation}

\begin{figure}
    \centering
    \includegraphics [width=0.8\linewidth]{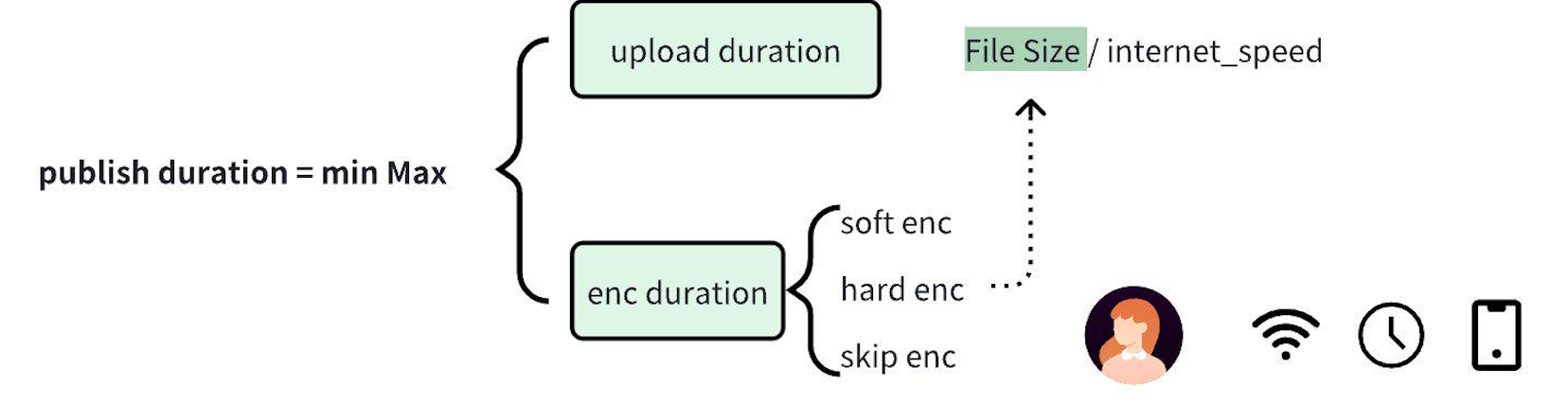}
    \caption{Smart Encoding Model }
    \label{fig:enc_model}
\end{figure}
Fig.~\ref{fig:enc_model} shows the idea of selecting encoding mode. Since uploading and synthesis encoding can be performed simultaneously, when synthesizing a video, processed part can be uploaded directly instead of waiting for the rest. User-perceived publishing duration $PublishDuration$ depends on maximum value of upload duration $UploadDuration$ and synthesis duration $EncodingDuration$, and $UploadDuration$ depends on file size brought by synthesis method $a_{enc} \in A_{enc}$. In practice, we can directly decide how to choose $a_{enc} \in A_{enc}$, estimate branches such as skip, soft, hard, and selection probability of subsequent parameters in each branch, so as to minimize the publishing time. 

Note encoding duration is corrected according to $\alpha_{u,i}$, that is, $EncodingDuration = Fix (EncodingDuration,\alpha_{u,i})$. High value videos are allowed to be encoded in a longer time. 
Although increasing $PublishDuration$ may reduce author satisfaction, the corresponding $ContentQuality$ improvement will bring benefits from consumers, thereby boosting the producers' enthusiasm, which will in turn improve author reward. It can still achieve an overall improvement. 

\noindent \textbf{Dynamic Encoding Parameters}

Encoding parameters have a great impact on upload experience. A common practice is to define preset configurations, but it is easy to ignore each user's expectation. Similar to above chapters, we first define all optional sets, establish a Decider function and integrate it into personalized feature solving, to seek a better balance between $ContentQuality$ and $PublishDuration$, see formula  \eqref{equ:publish_ltv_param}: 
\begin{equation}
\begin{aligned} \label{equ:publish_ltv_param}
\arg \max_{\pi (A_{param}|s)} & \Delta LTV (ContentQuality (\pi,s)|s) \\
& + \Delta LTV (PublishDuration (\pi,s)|s)
\end{aligned}
\end{equation}

Other elements are similar to those in previous formulas. The action selection $A_{param}$ includes various encoding-related parameters such as QP, FPS, HDR, Codec, Bitrate, Audio Channel, etc. Our idea is shown in Fig.~\ref{fig:enc_param}. Steps as follows: 
\begin{figure}
    \centering
    \includegraphics [width=1.0\linewidth]{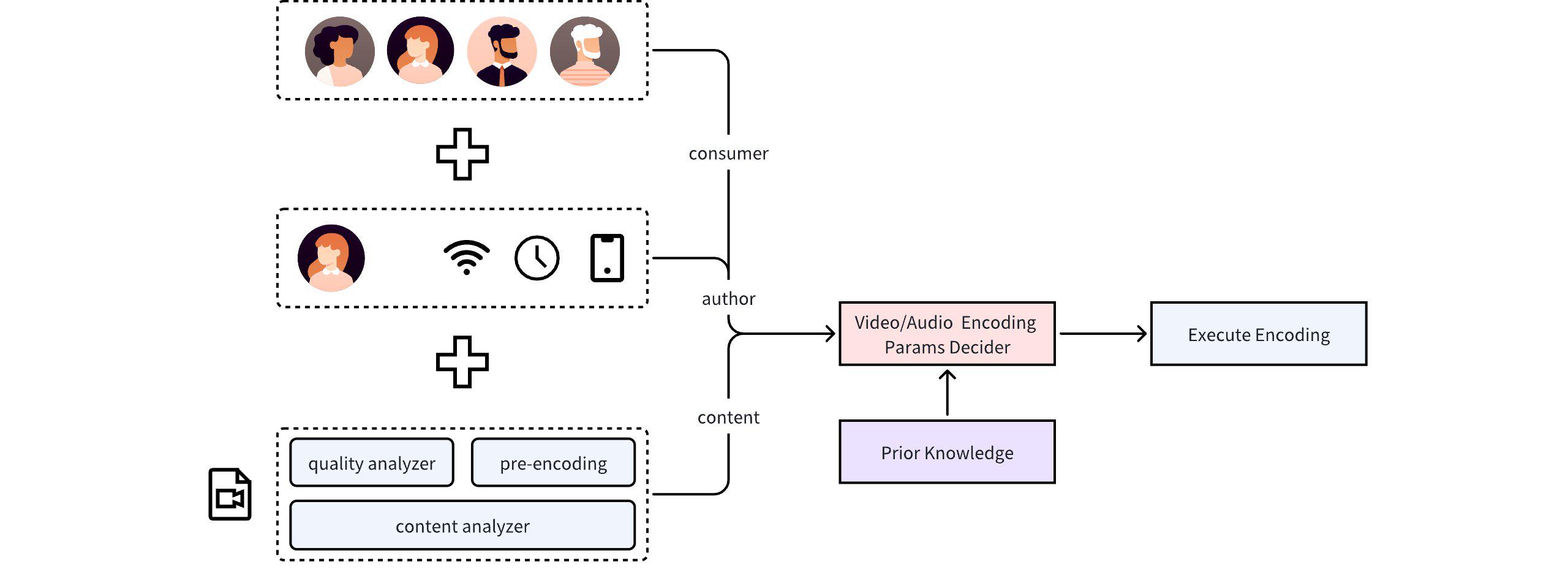}
    \caption{Dynamic Encoding Parameters}
    \label{fig:enc_param}
\end{figure}
\begin{enumerate} [leftmargin=*]
\item Analyze video content, including understanding the content, performing quality assessment and pre-encoding, etc. 
\item Estimate author's expectations, including preference for video quality, upload efficiency, video value, etc. 
\item Estimate the consumption value, as $\alpha_{u,i}$ mentioned above. 
\item Submit estimates to Decider to make decisions on encoding parameters for Video and Audio respectively. Prior knowledge (e.g. dynamically adjust parameters) needs to be introduced into the process for assistance. The specific decision can use rules or regression models, but a trade-off between accuracy of the decision algorithm and the additional computing overhead needs to be made. 
\end{enumerate}
\subsection{Meticulous Upload} \label{subsec:Meticulous_Upload}
In the process of uploading, there are also several treatment points, such as selecting a higher-quality node from all available ones, enabling chunk upload mode~\cite{Chunked} or streaming upload mode~\cite{Uploads}. Chunk upload is a common method, which is to divide the file into different sizes and upload them separately. Streaming upload divides the entire file into smaller ranges and sends them continuously.

\begin{figure}
    \centering
    \includegraphics [width=1.0\linewidth]{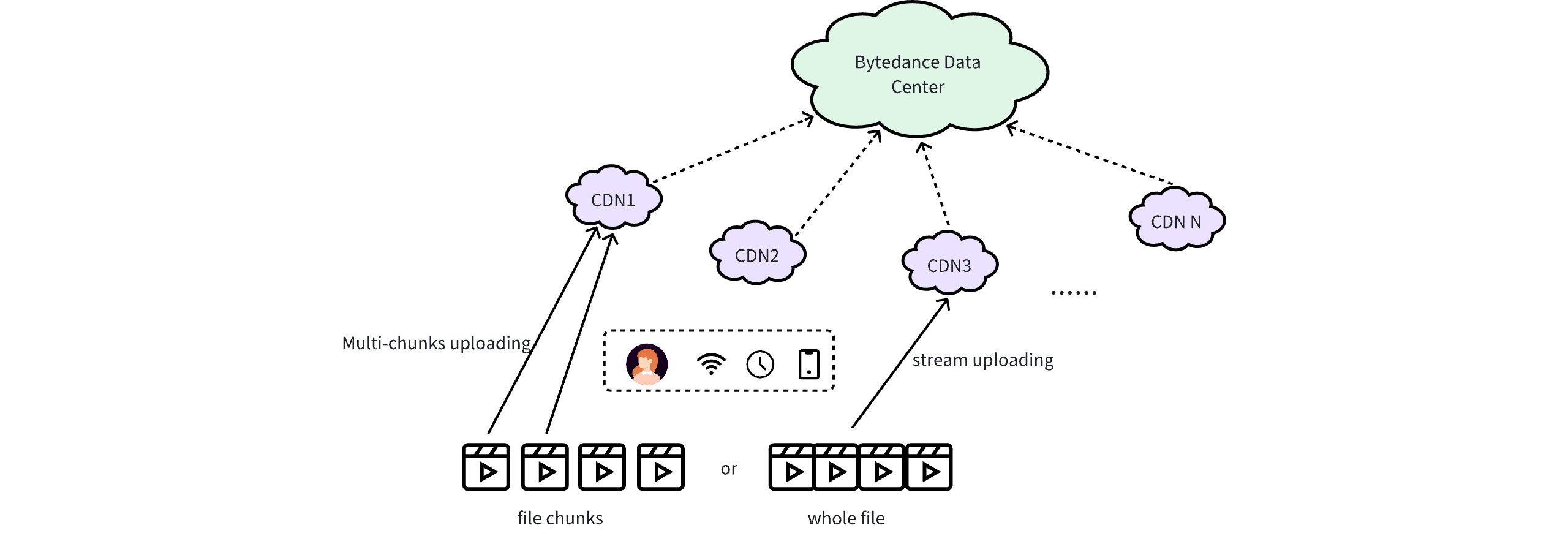}
    \caption{Brief of Upload Process}
    \label{fig:upload}
\end{figure}

The common upload process shown in Fig.~\ref{fig:upload}, we use $UploadDuration$ as optimization target, formula is \eqref{equ:publish_ltv_upload_method}, 
\begin{equation} \label{equ:publish_ltv_upload_method}
\begin{footnotesize}
\begin{aligned}
& \arg\max_{{\pi
 (a|s)}} 
  \Delta LTV (QoP (\pi,s))  \\   
& \approx \arg\min_{{\pi
 (a|s)}} \frac{1}{P} \sum\limits_i ^{chunks} UploadDuration (\pi,s,i)\times Repeat (i)  \\
& + ConnectDuration (\pi,s,i) \\  
\end{aligned}
\end{footnotesize}
\end{equation}

The duration of a video upload is mainly affected by connection duration $ConnectDuration$ and $UploadDuration$. Streaming upload mode can be regarded as a chunk upload with the number of chunks equal to 1. $Repeat (i)$ indicates the retry numbers, and $P$ indicates parallel chunk number. In formula, upload decision $a$ mainly includes selection of node quality, file chunk size, whether streaming upload or not, and chunk parallel granularity (Note: node quality selection strategy is similar to that described in chaper~\ref{subsec:scheduling}.), which will be discussed below. 

Main influencing factor of chunk upload is that if the shards too small, overall connection time will increase, but the failure rate of the chunks will decrease. If too large, opposite will happen. Meanwhile, chunk size is often difficult to adjust in time. Also, since chunk uploads usually use ordinary protocols, they are friendly to upload nodes and easy to use more complex technologies such as parallel and multi-node uploads. Streaming upload reduce retries through breakpoint-resume mechanism and reduce ack waiting time and response between shards, thereby reducing upload time. Advantage is transmission process control is flexible, and disadvantage is complicated fault-tolerant implementation and not easy to parallelize. So it requires personalized decision-making  (Upload Decider in Fig.~\ref{fig:ParameterDecision}) includes:
\begin{figure}
    \centering
    \includegraphics [width=1.0\linewidth]{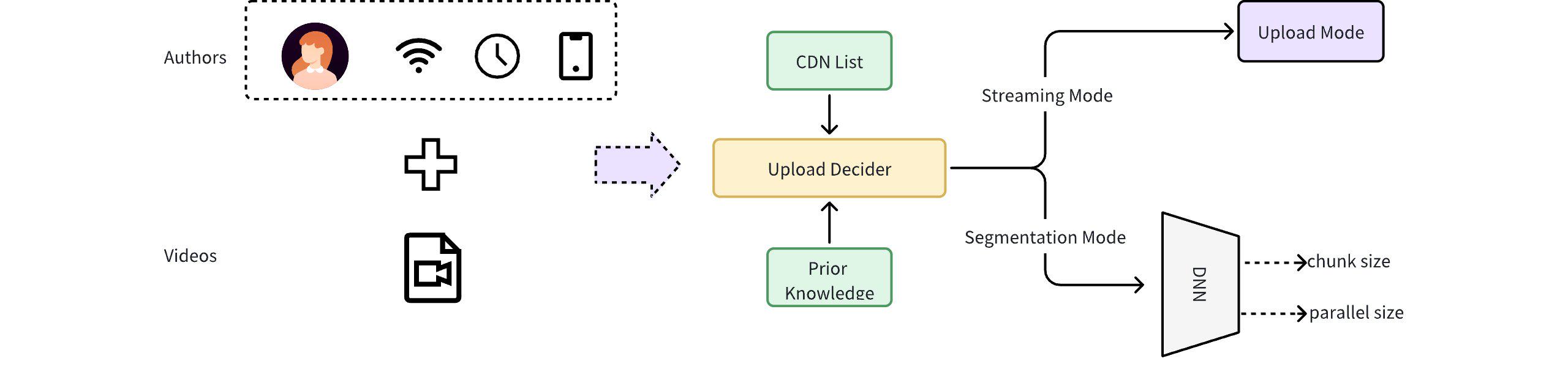}
    \caption{Personalized Upload Parameter Decision Process}
    \label{fig:ParameterDecision}
\end{figure}

\begin{enumerate} [leftmargin=*]

\item Determine chunk upload or streaming upload. 
\item If chunk mode selected, decide chunk size and parallel number, upload node, etc. 
\end{enumerate}

Besides, in order to further shorten publishing time, it can also consider using pre-publishing technology, before user click the "Publish" button, by encrypting and uploading the synthesized video in advance, so as to shorten the upload time perceived by the user after the actual action. Note, this technology should only be used with authorization, and purpose of encryption operation is to prevent anyone else from getting a non-final version. Here the optimization goal will be changed to \eqref{equ:publish_ltv_upload_method_encrypt}, where $EncryptDuration$ represents time required for encryption. If lead time $PreUploadLeadDuration$ of pre-publishing part larger, optimization achieved: 
\begin{equation} \label{equ:publish_ltv_upload_method_encrypt}
\begin{footnotesize}
    \begin{aligned}
\arg\min 
 & \{\max (0, UploadDuration+EncryptDuration \\
 & - PreUploadLeadDuration) \}
\end{aligned}
\end{footnotesize}
\end{equation}
Obviously, if user hesitates too long before posting, he can even achieve the effect of successful posting by clicking on it. However, this method also has obvious cost. The user may cancel the posting action or roll back, resulting in waste. Therefore, it is also necessary to estimate submission behaviors so that benefits brought by corresponding savings are greater than losses caused by prediction error. 
\subsection{Adaptive Priority} \label{subsec:Adaptive_Priority}
After clicking to publish, user often switch to other pages to continue browsing, or even switch to other apps. This will cause the upload process priority to be reduced or even killed, which leads to upload failure. A simple solution is to force increasing the upload process priority (for example, switching the process to the foreground). Although this method can improve success rate (Upload $QoP$), it may degrade other experiences (i.e., freeze on the consumer side, increase system power consumption, etc.), causing dissatisfaction. Therefore, how to more reasonably allocate process priorities has become a problem. As shown in Fig.~\ref{fig:upload_priority}:
\begin{figure}
    \centering
    \includegraphics [width=0.7\linewidth]{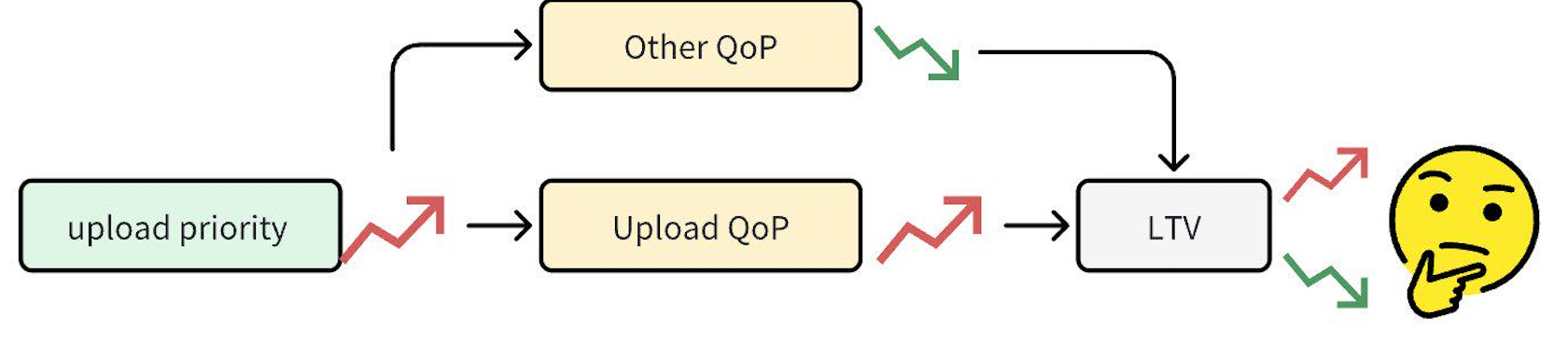}
    \caption{Upload Priority Decision}
    \label{fig:upload_priority}
\end{figure}

For process priority decision, we can define the optimization goal as maximizing the sum of $Profit$ of publish and consume (if switched to the background, $Profit$ of consume is 0), the formula is as \eqref{equ:publish_ltv_upload_method_uc}: 
\begin{equation} \label{equ:publish_ltv_upload_method_uc}
\begin{footnotesize}
\begin{aligned}
\arg&\max_{{\pi_{p}
 (a_{p}|s)}}
\overbrace{Profit (QoP ( \pi_{p},s))}^{publish} + 
\overbrace{Profit (QoP ( \pi_{c},s))}^{consume} \\ 
&s.t. \ \ Quota_p (a_p, s)+ Quota_c (a_c, s) \le MaxQuota \\ 
\end{aligned}
\end{footnotesize}
\end{equation}

Among them, $a_p$ is upload process priority, $\pi_p$ is upload process priority adjustment strategy, $a_c$ is consumer process priority, $\pi_c$ is consumer process priority adjustment strategy. $\pi_p$ and $\pi_c$ are independent strategies, and both need to satisfy the total resource $MaxQuota$ constraint. $Quota$ represents resource consumption, such as CPU, memory, bandwidth, etc. To simplify the computational complexity, we use a greedy strategy, that is, maximizing the $a_p$ priority while keeping $QoP_{consume}$ approximately unchanged, formula like  \eqref{equ:publish_ltv_upload_method_uc_greedy}, 
\begin{equation}
\begin{aligned} \label{equ:publish_ltv_upload_method_uc_greedy}
&\ \max\ a_p \ \ \ \ \ s.t.\  \Delta QoP (\pi_c|a_p, s) \gtrsim 0
\end{aligned}
\end{equation}

For example, when application is performing an upload action, user switches the app to background, then we need to transfer the background upload process to foreground (increase the priority) to ensure upload success rate to improve user experience. However, when user is working on other pages, upload priority should not change, and the synthesis or transmission should even be suspended to release resources to other processes. Although the failure rate of the submission itself may increase, user's immediate interests are better satisfied, then overall experience is improved.

\section{Results and Conclusion} \label{sec:conclusion}

Compared with traditional multimedia work, incremental benefits of the above system are as Table~\ref{table:profitAll}  (not including conventional work such as encoder upgrades, transmission protocol optimization, architecture modification, etc.), which is the incremental advantage brought by the system.
\begin{table*}
    \centering
    \caption{
    PROFIT OF PERSONALIZED PLAYBACK TECHNOLOGY IN 2019-2023.
    }\label{table:profitAll}
    \begin{tabular}{p{6cm}p{2cm}p{7cm}}

    \hline
        \textbf{Direction} & \textbf{Gain Type} & \textbf{Margin} \\
    \hline
    \multirow{3}{*}{Personalized Streaming \& Playback} & LT, ARPU & LT +x\%, PlayTime +x\%, Advv +x\%\\
    & Cost & -25\% \\
    & QoP & First Frame Duration -40\%, Rebuffer Rate per vv -10\% \\
    \hline
    \multirow{2}{*}{Personalized Quality Scheduling} & LT & LT +0.x\% \\
    & Cost & -5\% \\
    \hline
    \multirow{3}{*}{User-Item Aware Encoding} & LT, ARPU & PlayTime +0.x\%, Advv +x\% \\
    & Cost & -10\% \\
    & QoP & Rebuffer Rate per vv -10\% \\
    \hline
    \multirow{2}{*}{Foresight Based Publishing} & LT & Publish +x\% \\
    & QoP & Publish Success +3\%, Upload Duration -20\% \\
    \hline
    \end{tabular}

\end{table*}

From competitive perspective, ByteDance has gradually become tier 1 short video service provider in different markets since 2019, which indirectly confirms the success of personalized playback technology. 
In this article, we systematically created the cutting-edge and interdisciplinary field, clearly define a series of sub-problems in this technology, and provide practical solutions to some problems. In this new area, we believe that people should: 
\begin{itemize} [leftmargin=*]
    \item Establish a complete workflow, focusing on the use of precise assessment techniques like AB experiments to ensure that each technical project brings business profits. 
    \item Use personalized technical paradigms for optimization:
    \begin{itemize} [leftmargin=*]
    
      \item Expand the conventional $QoS$ indicator to a more comprehensive performance indicator group $QoP$.
      \item Directly model business indicator $Profit$, and build personalized intermediate indicators. 
      \item Expand the use of comprehensive rather than simplified User, Item, and Context features.
      \item Use deep neural networks, reinforcement learning, causal inference, operations research and other technologies that are aligned with the forefront of recommendation, computational advertising, etc., for decision method itself or for feature and portrait construction. 
    \end{itemize}
    \item With personalized decision-making as the core, re-evaluate and plan various tasks in multimedia area:
    \begin{itemize} [leftmargin=*]
      \item Redesign multimedia frameworks, player solutions, and service systems to carry complex decision-making control capabilities and expand decision actions. 
      \item Redesign codecs and pre- and post-processing components, video quality assessment methods, and expand personalized adjustment capabilities.
      \item Redesign network transmission methods, CDN functions, and even business negotiation strategies to expand personalized adjustment capabilities. 
    \end{itemize}
    \end{itemize}

Although this article takes short video on-demand playback-related technologies as an example, its ideas are also fully applicable to work in other multimedia fields such as live broadcasting, to focus on additional real-time requirements in indicators and topology. On top of this, live system should also be abstracted as an optimization problem oriented to business goals, to drive evolution in each channel. For another example, the optimization scope discussed in this article is mainly concentrated within application, but various related and derived problems can also be transformed in the same way.

The widespread development of multimedia technology has been nearly 30 years, serving billions of users. Although the work based on encoding and video processing, transmission protocols, multimedia frameworks, elastic services and other buckets has been evolving, the difficulty of improvement is increasing continuously, and the marginal value is downing. However, it does not mean that the field of technology is dying. The personalized playback technology created by us has changed the boundaries and paradigm of this area. It can more imaginatively accommodate the development of various technologies(including LLM/AIGC) to integrate with traditional multimedia technologies, and can expand a new unlimited ceiling, through which billions of users can better enjoy their life improvement brought by multimedia. 

\section{Acknowledgments}
\noindent We are grateful to Xingrun Chen, Bing Yan, Mu Zhu, Mengjie Li for early attempt, also 
to Jing Ye, Yizhu Zhao, Quantong Chen, Rui Zhao and Siyi Gu for insightful data analysis, to Hui Wang, Zexin Luo, Min Wang, Chao Ma, Qing Huang, Kunzhi Gui, Mingkui Liu, Heng Liu, Linhui Sun, Ning Hao, Louyun Li, Ruying Hong, Fangshun Ge, Ti Gong, Zhengting Li, Moki Zhang, Junlin Li, Tao Zhang, Jianzhao Liu, Mingming Shen, Hanbang Liang for important implementation, to Ye Huang, Junhui Cui for wonderful tools support. 

~\\
Specially grateful to Han Li, Yuming Liang, Yue Wang \& Mingfei Hao for their visionary support.

\clearpage
\bibliographystyle{IEEEtran}
\bibliography{custom}
\end{document}